\documentclass[final,3p,times,twocolumn]{elsarticle}
\usepackage{amssymb, lineno, hyperref, subcaption}
\usepackage{pdfpages}
\journal{Journal of Alloys and Compounds}

\begin{document}

\begin{frontmatter}

\title{Structural and magnetic study of PrMn$_{1-x}$Fe$_x$O$_3$ compounds}

\author[addr1]{M. Mihalik jr.\corref{cor1}}
\cortext[cor1]{Corresponding author} 
\ead{matmihalik@saske.sk}
\author[addr2]{Z. Jagli\v ci\'c}
\author[addr3]{M. Fitta}
\author[addr1]{V. Kave\v cansk\'y}
\author[addr1]{K. Csach}
\author[addr3]{A. Budziak}
\author[addr4]{J. Brian\v cin}
\author[addr1]{M. Zentkov\'a}
\author[addr1]{M. Mihalik}
\address[addr1]{Institute of Experimental Physics Slovak Academy of Sciences, Watsonova 47, 040 01 Ko\v sice, Slovak Republic}
\address[addr2]{Institute of Mathematics, Physics and Mechanics and Faculty of Civil and Geodetic Engineering, University of Ljubljana, Slovenia}
\address[addr3]{Institute of Nuclear Physics Polish Academy of Sciences, Radzikowskiego 152, 31-342 Krak\'ow, Poland}
\address[addr4]{Institute of Geotechnics SAS, Watsonova 45, 043 53 Ko\v sice, Slovak Republic}

%
%
\begin{abstract}
The structural and magnetic study of PrMn$_{1-x}$Fe$_x$O$_3$ $(0 \leq x \leq 1)$ substitutional orthorhombical system, was performed on polycrystalline $(x \leq 0.5)$ as well as on the single crystalline $(0.5 \leq x)$ samples. In this system the orthorhombic – to orthorhombic structural phase transition takes place at high temperatures for $x \leq 0.2$ due to long range ordering of $3d_{3x^2-r^2}$ and $3d_{3y^2-r^2}$ Mn orbitals in Jahn-Teller distorted structure. We have found the Jahn-Teller distortion still contributes to orthorhombic distortion of the crystal lattice for samples with $x \leq 0.6$ at room temperature.  Our investigation of the magnetic phase transition with respect to chemical composition revealed the crossover region $0.4 \leq x \leq 0.5$ in which both Fe and Mn long range magnetic ordering exist simultaneously and the magnetic ordering of Mn sublattice is induced by already ordered Fe sublattice. The spin reorientation magnetic phase transitions are typical features for samples with $0.6 \leq x$. We have not found any traces of magnetic ordering of Pr ions at temperatures higher than 2 K.  These results enable us to construct complex structural and magnetic phase diagram for PrMn$_{1-x}$Fe$_x$O$_3$ system.
\end{abstract}

\begin{keyword}
A. oxide materials \sep C. phase diagrams \sep D. calorimetry \sep D. magnetic measurements \sep D. X-ray diffraction
\end{keyword}

\end{frontmatter}
%
%

\section{Introduction}
$RE$MnO$_3$ and $RE$FeO$_3$ compounds ($RE$ = Rare earth) have been extensively studied due to their high application potential in the field of hyperthermia (especially LaMnO$_3$ doped on La site \cite{melnikov2009, bride2016}), magnetocaloric effect at room temperature \cite{hien2002}, other magnetic applications, since $RE$FeO$_3$ compounds have ordering temperature well above room temperature \cite{bertaut1967, koehler1960, pinto1972}, but also as cathodes for solid oxide fuel cells \cite{huang2002, ran2005}. Among interesting phenomena in these compounds are for example multiferroicity ($RE$MnO$_3$ compounds with $RE$ element heavier than Gd \cite{kuwahara2005, goto2005}), interplay of the magnetism of two different magnetic sublattices (rare earth sublattice and Mn, or Fe sublattice) and possibility to tune the magnetism in these sublattices by doping \cite{ivanov2003, mihalik2013, ganeshraj2010}. Worth of note is also the possibility of studying the magnetic superexchange interactions in these materials, since several systems can be obtained in at least two different crystallographic modifications \cite{trukhanov2007, pena2002, antonak2014, markovich2005, guarin2012}. Since Mn$ ^{3+}$ is Jahn-Teller (JT) active ion, the degeneracy of manganese e$_{\rm g}$  orbital in Mn-based compounds can be removed by a cooperative JT distortion below transition temperature T$_{\rm JT}$ and there can be also observed the long range ordering of $3d_{3x^2-r^2}$ and $3d_{3y^2-r^2}$  orbitals below temperature $T ^*$ \cite{maris2004, zhou2003}.

The PrMnO$_3$ parent compound orders antiferromagnetically at $T_{\rm N}$ = 99 K \cite{hemberger2004}, where only the Mn$^{3+}$ sublattice orders into C$_{\rm x}$F$_{y}$ magnetic structure \cite{baran2013}. In the second parent compound PrFeO$_3$, the Fe$^{3+}$ sublattice orders antiferromagnetically well above the room temperatures \cite{pinto1972} and the magnetization vs. temperature data exhibit an anomaly at $T < 50$ K \cite{wang2013}. The recent work of Sultan et al. \cite{sultan2014}, who performed dielectric measurements on PrMn$_{1-x}$Fe$_x$O$_3$ $(0.5 \leq x \leq 1)$, indicates the anomalies in the real part of dielectric permittivity for temperatures lower than room temperatures and for $x \leq 0.7$. Recently, Ganeshraj et al. \cite{ganeshraj2010} demonstrated that the compound with $x$ = 0.5 orders ferromagnetically and the Curie temperature is less than 200 K.
 
Our work was motivated by the fact that the switching between ferromagnetic ordering for $x$ = 0.5 and antiferromagnetic ordering for $x$ = 0 and 1 remains puzzling problem. Another impulse for our research was that while many papers deal with substitution on Pr site \cite{huang2002, markovich2005, dyakonov2006, niebieskikwiat2005, zhang2009, rossler2011}, only small number is devoted to Mn-Fe substitution \cite{ganeshraj2010, sultan2014}. Since PrMnO$_3$ orders magnetically below 100 K and PrFeO$_3$ orders at very high temperature, one can expect that the substitution of Mn for Fe can tune magnetic ordering temperature of PrMn$_{1-x}$Fe$_x$O$_3$ in the temperature range around room temperature. Such a tuning can have application potential in magnetic switching, hyperthermia, or magnetic cooling applications.  All these facts motivated us to prepare the PrMn$_{1-x}$Fe$_x$O$_3$ substitutional solid solution and study the evolution of magnetism and crystal structure in this system with respect to chemical composition. 
%
%
\section{Sample preparation and experimental details}
%
%
\begin{figure}[t]
\begin{center}

\includegraphics[angle=0,width=0.47\textwidth]{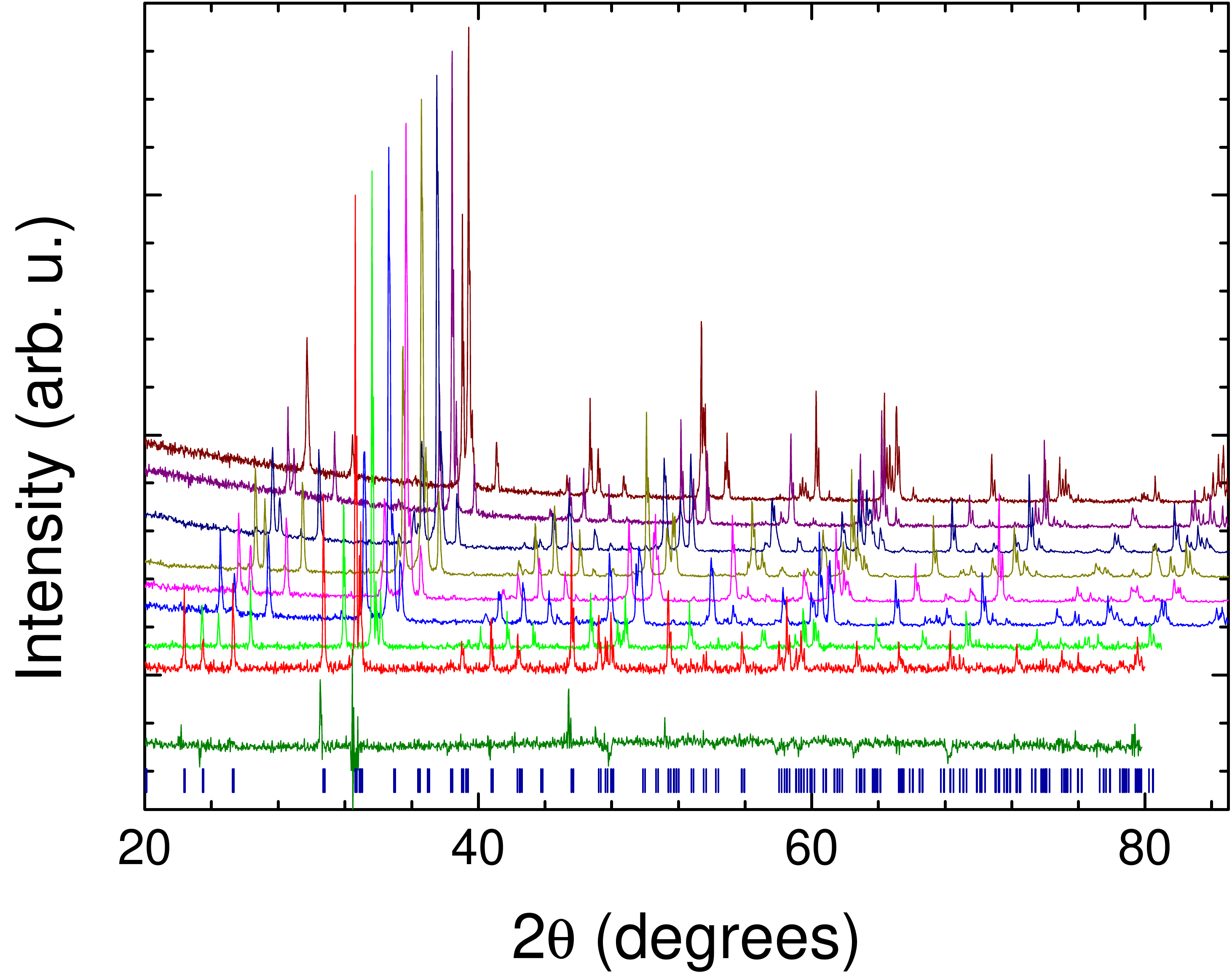}
\pdfcompresslevel=9

\caption{(Color online) The meassured XRPD patterns for (from down to up) $x$ = 0; 0.1; 0.2; 0.3; 0.4; 0.5; 0.6 and 1. The line below the spectra visualizes the difference between the experimental data obtained for concentration $x$ = 0 and Rietveld--refined crystalographic structure. For the Rietveld refinement the  PrMnO$_3$ crystal structure as presented by Baran et al. \cite{baran2013} was used as a starting model. The ticks below the lines represent the calculated crystallographic positions for PrMnO$_3$ parent compound. \label{fig0}}
\end{center}
\end{figure}
%
%
%
%
\begin{figure}[t]
\begin{center}

\includegraphics[angle=0,width=0.47\textwidth]{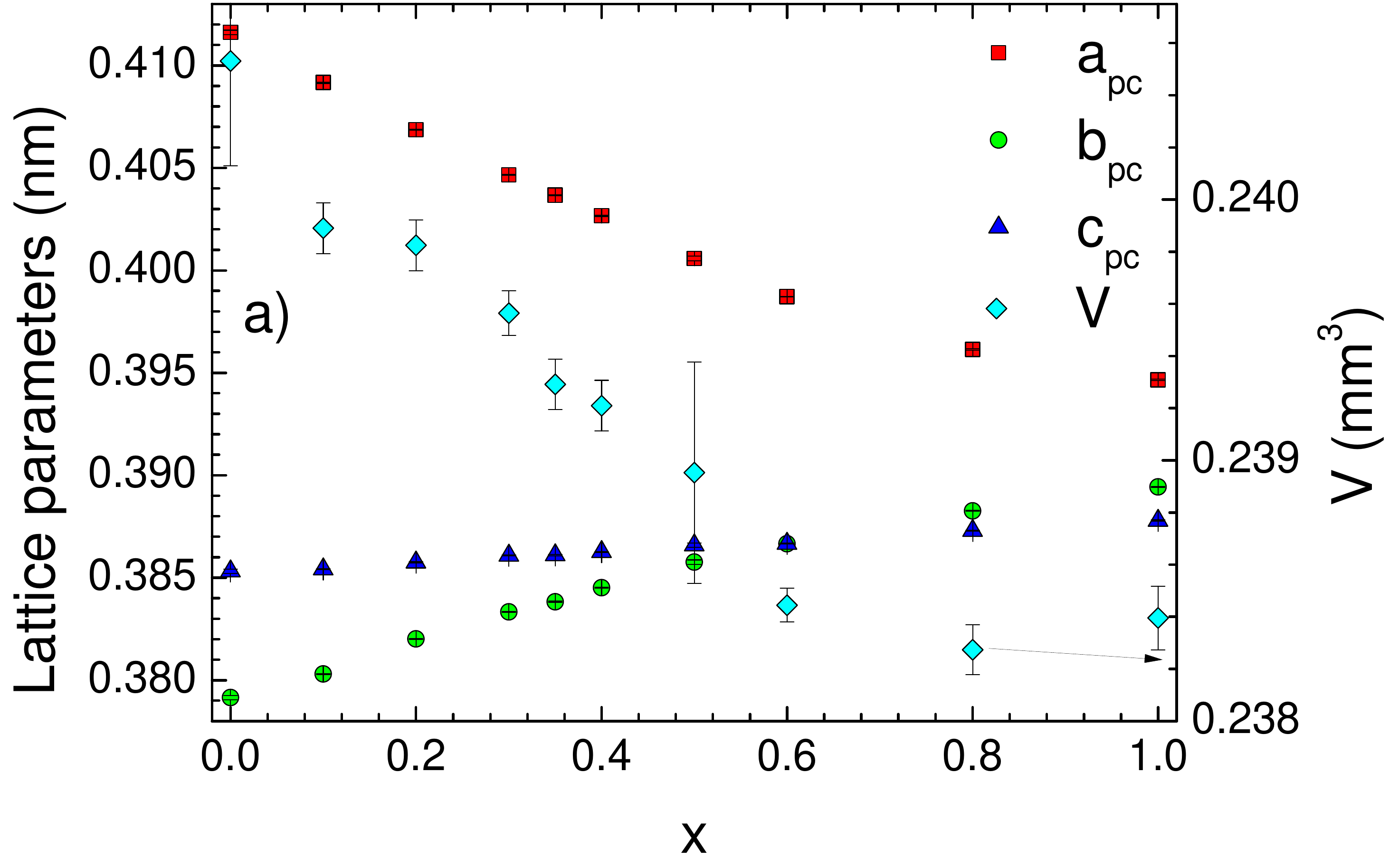}
\pdfcompresslevel=9

\includegraphics[angle=0,width=0.42\textwidth]{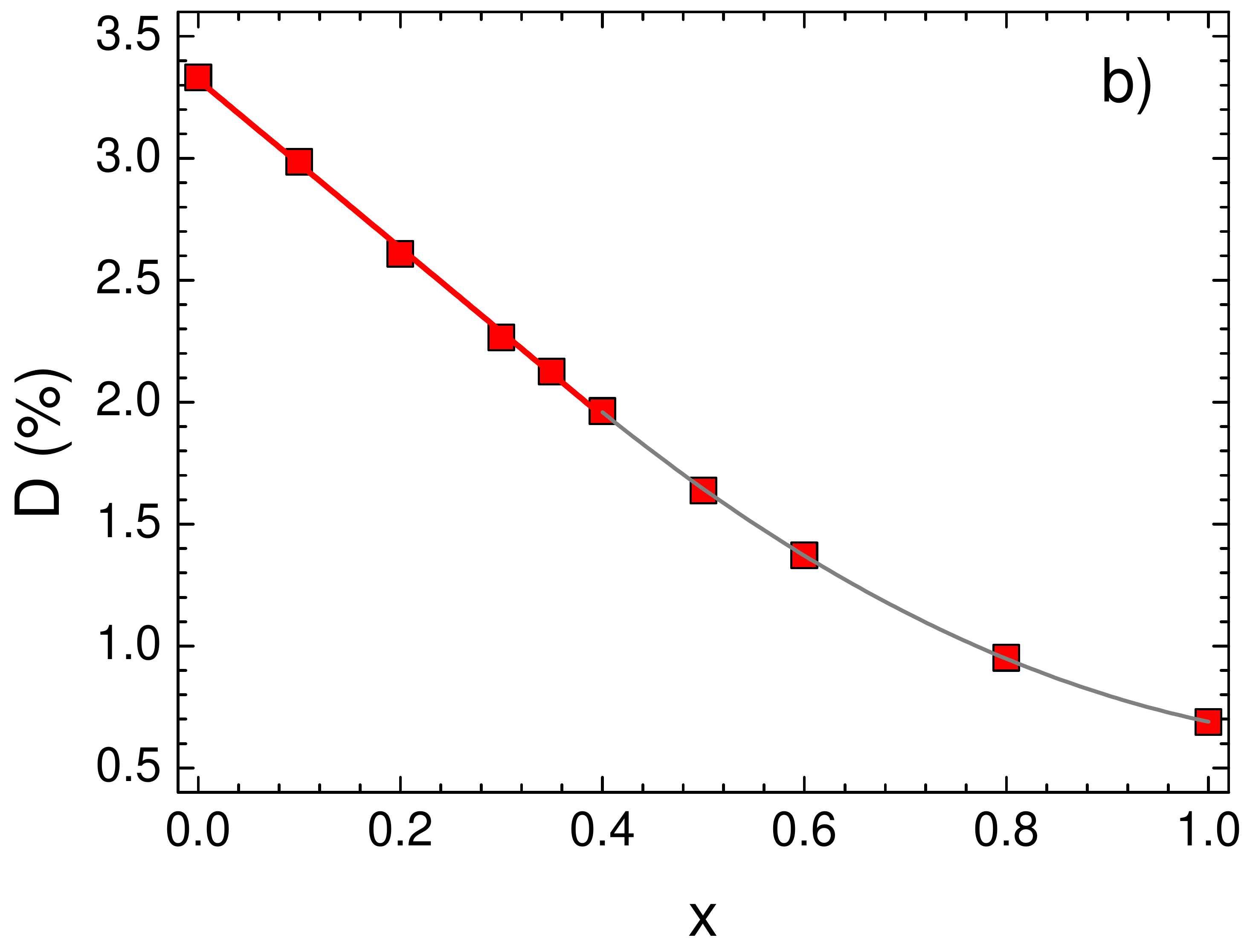}
\pdfcompresslevel=9

\caption{(Color online) The concentration evolution of a) the pseudocubic lattice parameters and b) the lattice distortion index at room temperature decreases linearly down to x = 0.4. \label{fig1}}
\end{center}
\end{figure}
%
%
Samples of PrMn$_{1-x}$Fe$_x$O$_{3+\delta}$ $(0 \leq x \leq 1)$ were prepared by floating zone method in 4-mirror optical furnace (type: FZ-T-4000 from Crystal Systems corporation). As starting materials we used MnO$_2$ (purity 3N, supplier: Alpha Aesar), Pr$_6$O$_{11}$ (purity 3N, supplier: Sigma Aldrich) and Fe$_2$O$_3$ (purity 2N, supplier: Sigma Aldrich). The starting materials were mixed in a Pr:Mn:Fe stoichiometric ratio as intended for the final compound. Subsequently the powders were cold pressed into rods and sintered at 1100 $^\circ$C for 12 to 14 hours in air. The parameters of crystal growth were as follows: rotation of both, upper and lower shaft 15 to 30 rpm; pulling speed 7 mm/h; feeding speed 6 mm/h and the crystal growth was performed in air atmosphere. The quality of grown rods was checked by Laue method using tungsten X-ray white radiation and image plate detector. For concentrations $x < 0.5$ we were able to prepare only the polycrystalline samples, however, for compositions $x = 0.5$; 0.6; 0.8 and 1 we have succeeded with the growth of single crystalline samples.  

X-ray powder diffraction (XRPD) measurements were performed on X'PERT PRO Panalytical diffractometer in standard Bragg - Brentano geometry with fixed slits and as a source we used Cu K$_{\alpha_1,\alpha_2}$ doublet radiation ($\lambda$K$_{\alpha_1}$ = 0.154060 nm, $\lambda$K$_{\alpha_2}$ = 0.1544430 nm) and Rigaku Ultima IV diffractometer in Bragg Brentano configuration. For the temperature evolution of the crystallographic parameters we have used HTK1200N (Anton Paar) High-Temperature Oven-Chamber temperature cell with air atmosphere for Ultima IV and X'PERT PRO diffractometers, and TTK450 (Anton Paar) Low-Temperature Chamber  for X'PERT PRO. Obtained data were processed using Le Baill and Rietveld method implemented in program FullProf \cite{carvajal1993}. EDX analysis was performed on scanning electron microscope (SEM) Mira III FE (produced by Tescan), which was equipped with analyzer PentaFET Precision (produced by Oxford Instruments). For all concentrations we have examined two parts of the grown ingot – beginning and end of grown ingot. XRPD investigation at room temperature confirmed that all samples are single phase within the precision of this experimental technique. SEM and EDX analysis confirmed that both parts of the studied ingots are free of any inclusions, no concentration gradient between the beginning and the end of the crystal was found, and Nd : Mn : Fe ratio was equal to the expected one within the experimental error

Magnetization (M) and AC susceptibility (χAC) measurements were performed on the SQUID magnetometers (MPMS-XL-5 and MPMS3) from Quantum Design in the temperature range from 2 to 300 K and in applied magnetic fields up to 5 T. In the case of samples with $x < 0.5$ we have used powders, which were freshly powdered prior the experiment and fixed by paramagnetic glue into the paramagnetic capsules. The usual mass of the samples varied between 20 and 30 mg (DC magnetization) and 300 -- 400 mg (AC susceptibility). In the case of Fe concentration $x \geq 0.5$ we have used single crystals with mass between 10 and 50 mg and the samples were fixed by GE varnish glue on glass, or brass holders. PPMS (Quantum Design) apparatus with oven option was used for the high temperature magnetic measurements up to 1000 K. In this case the samples were fixed by Zirconia cement.

Molar heat capacity $(C)$ data for $x < 0.5$ were measured by relaxation method on bulk, highly textured samples directly cleaved from the grown ingot. The sample’s mass was between 7 and 12 mg. The measurements were performed on PPMS apparatus in the temperature range 2 -- 220 K. Differential scanning calorimetry (DSC) was performed on Q2000 (TA Instruments) apparatus in temperature range 240 -- 400 K and nitrogen atmosphere. Samples with masses of about 40 mg pressed into aluminum pans were used for these experiments. The rate of heating and cooling was equal; 10 K/min. Termogravimetry (TG) and Differential thermal analysis (DTA) were simultaneously performed by Setaram-SETSYS16 apparatus in flowing air (100 ml/min) atmosphere and with heating rate 10 K/min on bulk pieces of samples, which were housed in alumina crucibles.
%
%
\section{Results} 
\subsection{ Crystallography}
%
%
\begin{figure}[t]
\begin{center}
\includegraphics[angle=0,width=0.47\textwidth]{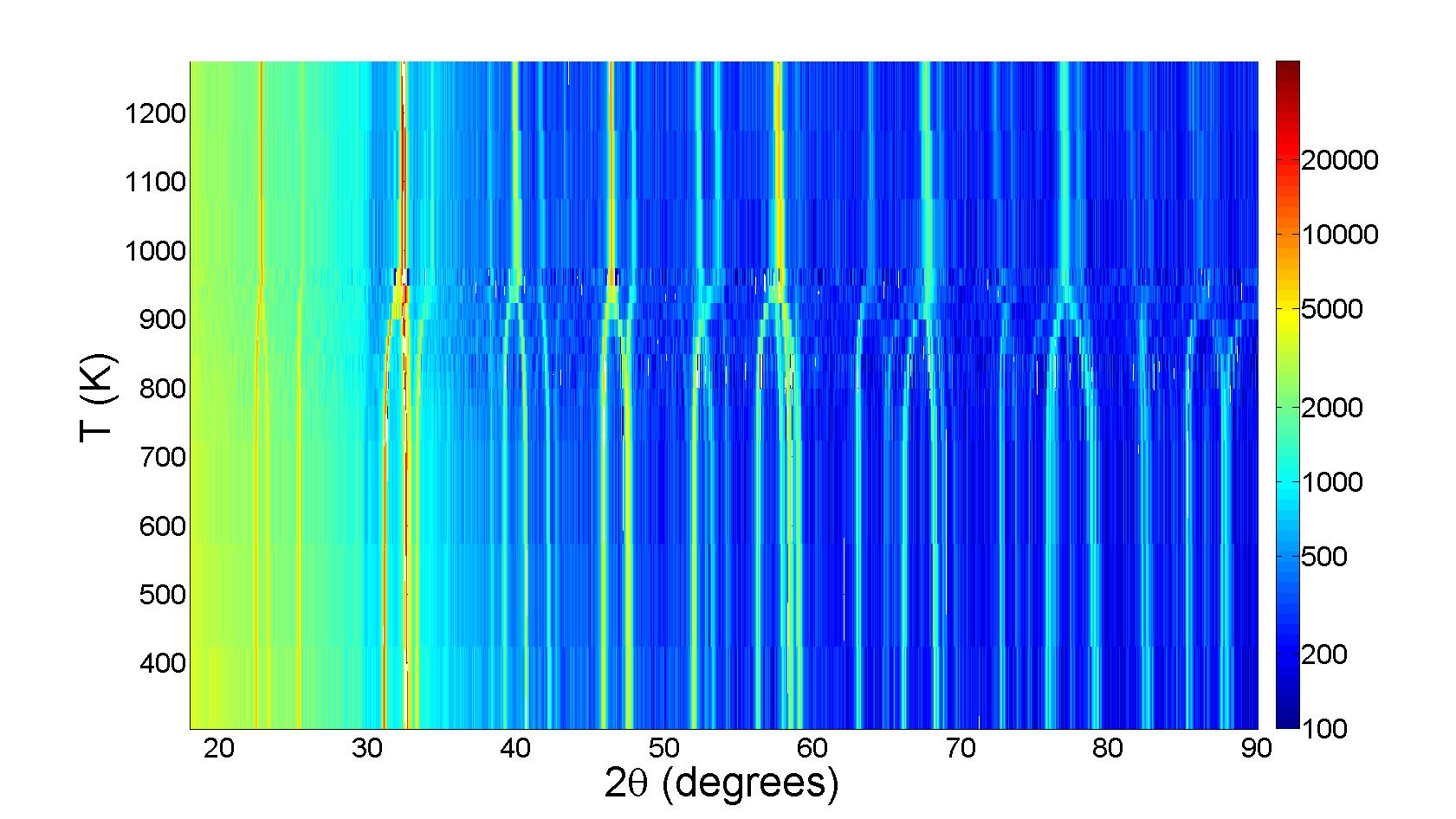}\\
\pdfcompresslevel=9
\includegraphics[angle=0,width=0.47\textwidth]{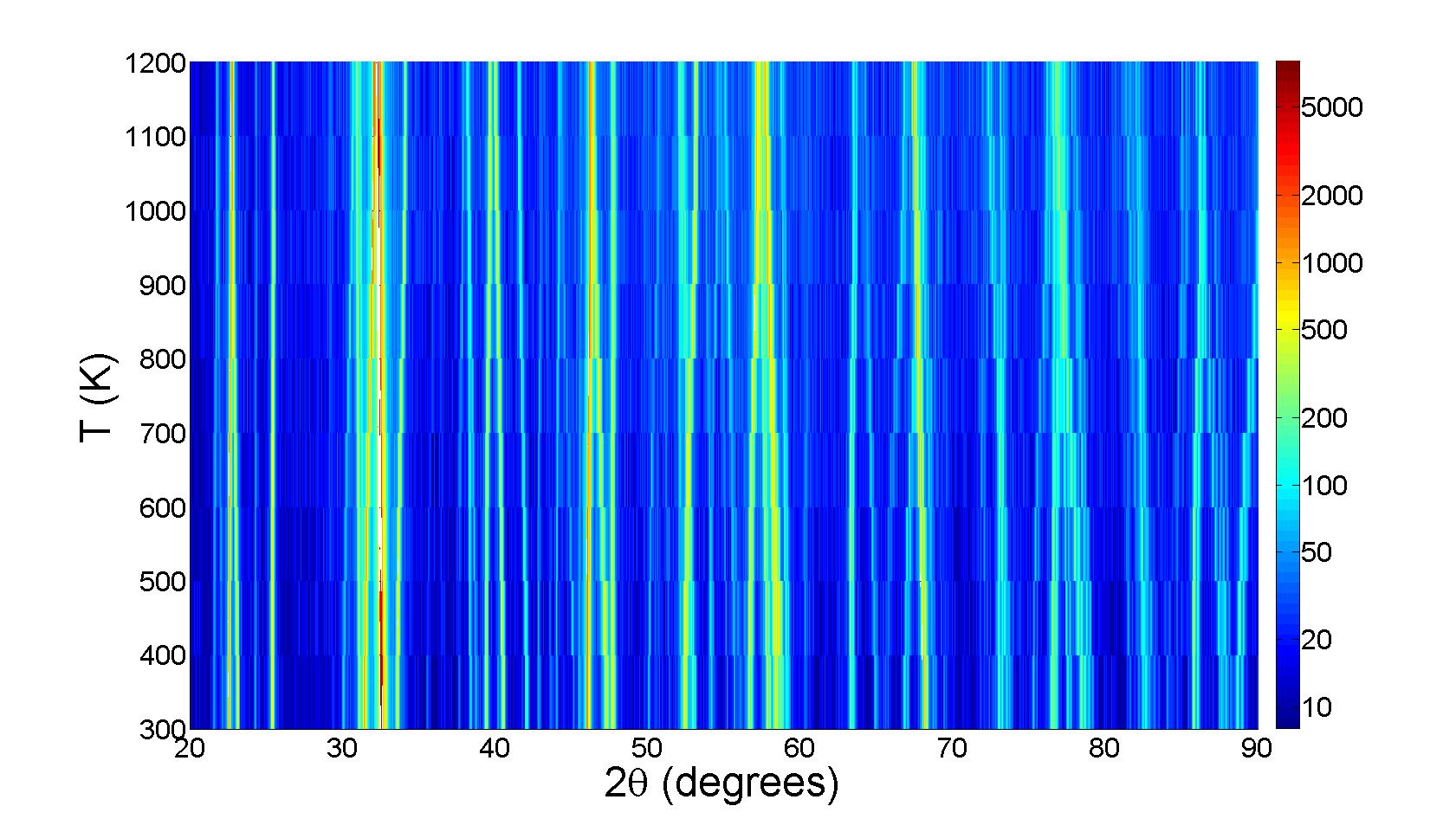}
\pdfcompresslevel=9
\caption{(Color online) The temperature evolution of XRPD patterns for PrMn$_{0.8}$Fe$_{0.2}$O$_3$ (upper panel) and PrMn$_{0.6}$Fe$_{0.4}$O$_3$ (lower panel). Temperature evolutions for the other concentrations are presented in Supplementary online material, section 1. \label{fig2}}
\end{center}
\end{figure}
%
%
%
%
\begin{figure*}[t]
\begin{center}
\includegraphics[angle=0,width=0.95\textwidth]{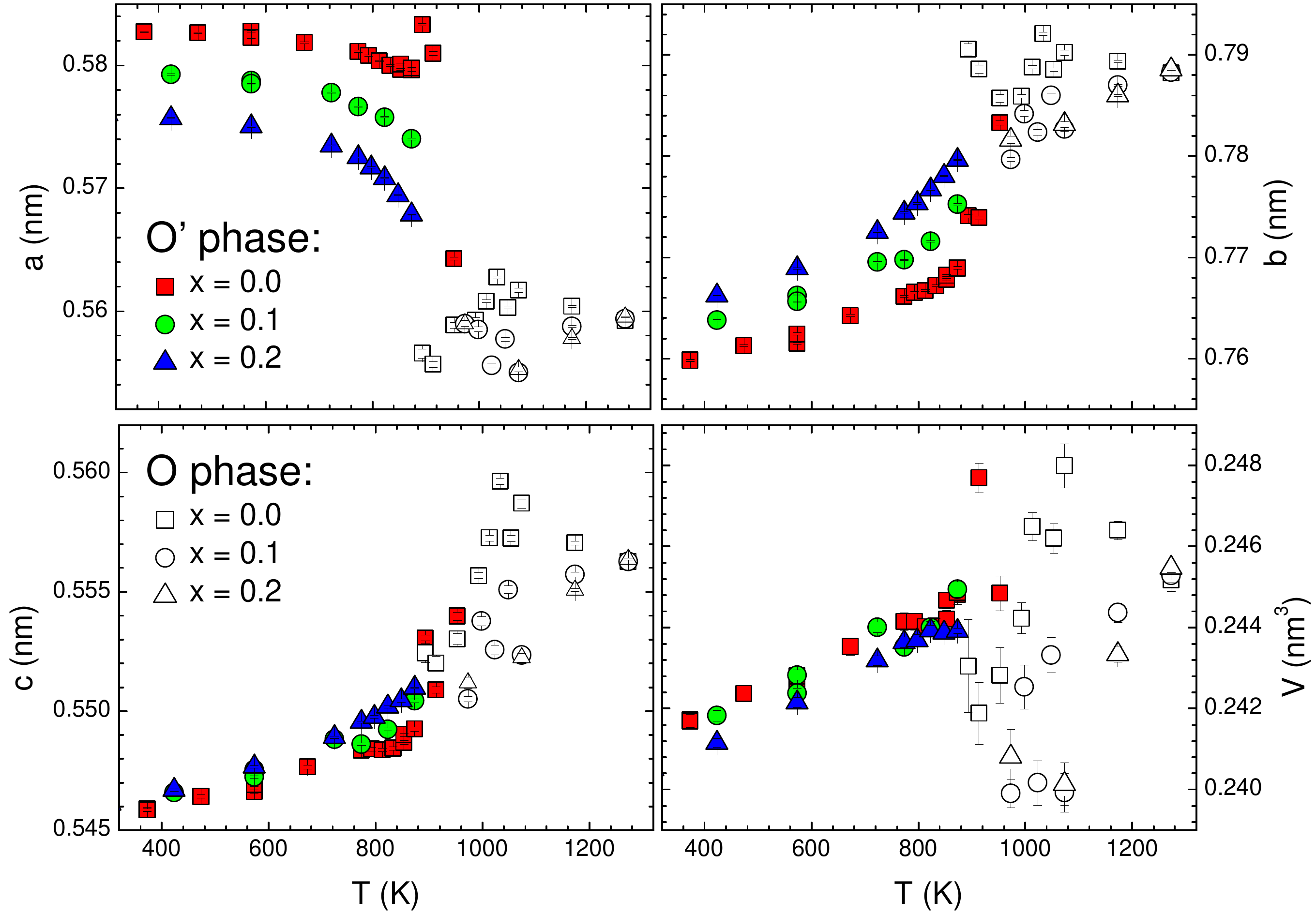}
\pdfcompresslevel=9
\caption{(Color online) Temperature evolution of crystallographic parameters and calculated volume for both, O' and O phases for PrMnO$_3$, PrMn$_{0.9}$Fe$_{0.1}$O$_3$ and PrMn$_{0.8}$Fe$_{0.2}$O$_3$ compounds. \label{fig3}}
\end{center}
\end{figure*}
%
%

The structure of rare-earth perovskite oxides, $RE$MnO$_3$ and $RE$FeO$_3$, can be described as GdFeO$_3$-type orthorhombic distortion from the original, cubic CaTiO$_3$ perovskite, or, using Glazer notation \cite{glazer1972}, as $a a^-b^+a^-$ tilt.  Mn$^{3+}$ ions are Jahn-Teller (JT) active ions, but Fe$^{3+}$ ions are not JT active. Since both ions have practically the same Shanon radii (in high spin state, which is expected in this case, both have radius 0.785 \AA) \cite{shannon1976} one should expect that the Mn-Fe substitution will directly probe the JT effect in the PrMn$_{1-x}$Fe$_x$O$_3$ system. Parent compounds $RE$MnO$_3$ and $RE$FeO$_3$ have been already extensively studied from the crystallographic point of view and the results can be found elsewhere \cite{maris2004, alonso2000, lufaso2004, marezio1970}. The reference of the pseudo-cubic lattice parameters which are defined as follows:

\begin{equation}
a_{pc}=\frac{a}{\sqrt{2}}; \  \ b_{pc}=\frac{b}{2} ; \  \ c_{pc}=\frac{c}{\sqrt{2}}                 
\end{equation}

($a$, $b$ and $c$ are the true, orthorhombic lattice parameters) is usually very convenient for study of the JT distortion and the orthorhombic distortion. The lattice distortion index ($D$) can be expressed by the formula \cite{lufaso2004}:

\begin{equation}
D=\frac{1}{3} \sum^3_{n=1}100\frac{\left | a_n-\left <a \right>\right |}{\left <a \right >}
\end{equation}
where $a_n$ are pseudo-cubic lattice parameters and $\left <a \right >$ is the mean value of $a_n$. In principle, $D$ is the measure of the deviation from the cubic lattice. 
The concentration evolution of the pseudo-cubic lattice parameters (Fig. \ref{fig1}a) clearly shows two regimes: $a_{pc} > c_{pc} > b_{pc}$ for $x < 0.6$, which can be typically found in perovskites whose distortions are dominated by both tilting of the octahedrons and JT distortion and region $a_{pc} > b_{pc} > c_{pc}$ for $x > 0.6$, where the distortion of the crystal lattice is driven by tilting of the octahedrons. The distortion index $D$ determined at room temperature monotonously decreases with increasing iron concentration, which means that the iron substitution brings the structure closer to original, perovskite structure (Fig. \ref{fig1}b).  Similar conclusions can be extracted from previous research published by Sultan et al. \cite{sultan2014}, but in this case the relation $a_{pc} > b_{pc} > c_{pc}$ holds already for $x \geq 0.5$. Since Sultan et al. \cite{sultan2014} prepared samples by solid state reaction, the difference is probably caused by the different preparation route and thence the different amount of Mn$^{4+}$ ions which implies the different oxygen off-stoichiometry $\delta$. The similar effect of crossing $b_{pc}$ and $a_{pc}$ with iron concentration was also observed for DyMn$_{1-x}$Fe$_x$O$_3$ system for concentration of about $x$ = 0.4 \cite{chiang2011}. 

The temperature evolution of XRPD patterns for $x \leq 0.2$ revealed the structural phase transition at temperatures around 900 - 1000 K (see Fig. \ref{fig2} and Supplementary online material, section 1). Two different crystallographic phases coexist in this temperature region which implies that the structural phase transition is clearly of the first order. Possible model structures for higher temperatures were found with help of program package SPuDS \cite{lufaso2001}, which incorporates into its calculations all distortions of the perovskite structure as published by Glazer \cite{glazer1972}. The best Rietveld fit for $x \leq 0.2$ concentrations was obtained again for the orthorhombic structure with space group $Pnma$, however, with different crystallographic parameters (Fig. \ref{fig3}). In this high temperature phase $a$-axis length approaches the $c$-axis length. Similar structural change was previously observed for LaMnO$_3$ at temperature varying from 600 K \cite{norby1995} to 750 K \cite{carvajal1998} depending on the content of Mn$^{4+}$ ions, for PrMnO$_3$ at temperature approximately 950 K \cite{pollert1982} and for NdMnO$_3$ in temperature range 1000 – 1200 K \cite{maris2004}. Since all previously mentioned authors adopt notation O' for low temperature phase and O for higher temperature phase, we adopt the same notation. In our case, in O phase the equation $a_{pc} > b_{pc} > c_{pc}$ is fulfilled, while in O' phase the equation $a_{pc} > c_{pc} > b_{pc}$ is fulfilled (see Supplementary online material). It means that O phase is not JT distorted, while O' is JT distorted, and that is why we assign the O -- O' structural phase transition to the onset of Jahn-Teller distortion and we denote the corresponding temperature as $T_{\rm JT}$. In the temperature region approximately 60 K below $T_{\rm JT}$ it is impossible to fit the XRPD spectra by single crystallographic structure. Instead of this, one has to use the crystallographic model, which consists of two phases: O and O'. This effect can be ascribed to the ordering of $3d_{3x^2-r^2}$ and $3d_{3y^2-r^2}$ orbitals, which does not take place at exact temperature, but in the whole temperature interval $T^* < T < T_{\rm JT}$. Below $T^*$ we have observed only the O' phase, which is orbitally-ordered phase. Note that the observation of the orbital ordering was already reported for several pure $RE$MnO$_3$ compounds \cite{ zhou2003, qiu2005, murakami1998, chatterji2003}, but not for doped compounds.

\begin{figure}[t]
\begin{center}
\includegraphics[angle=0,width=0.47\textwidth]{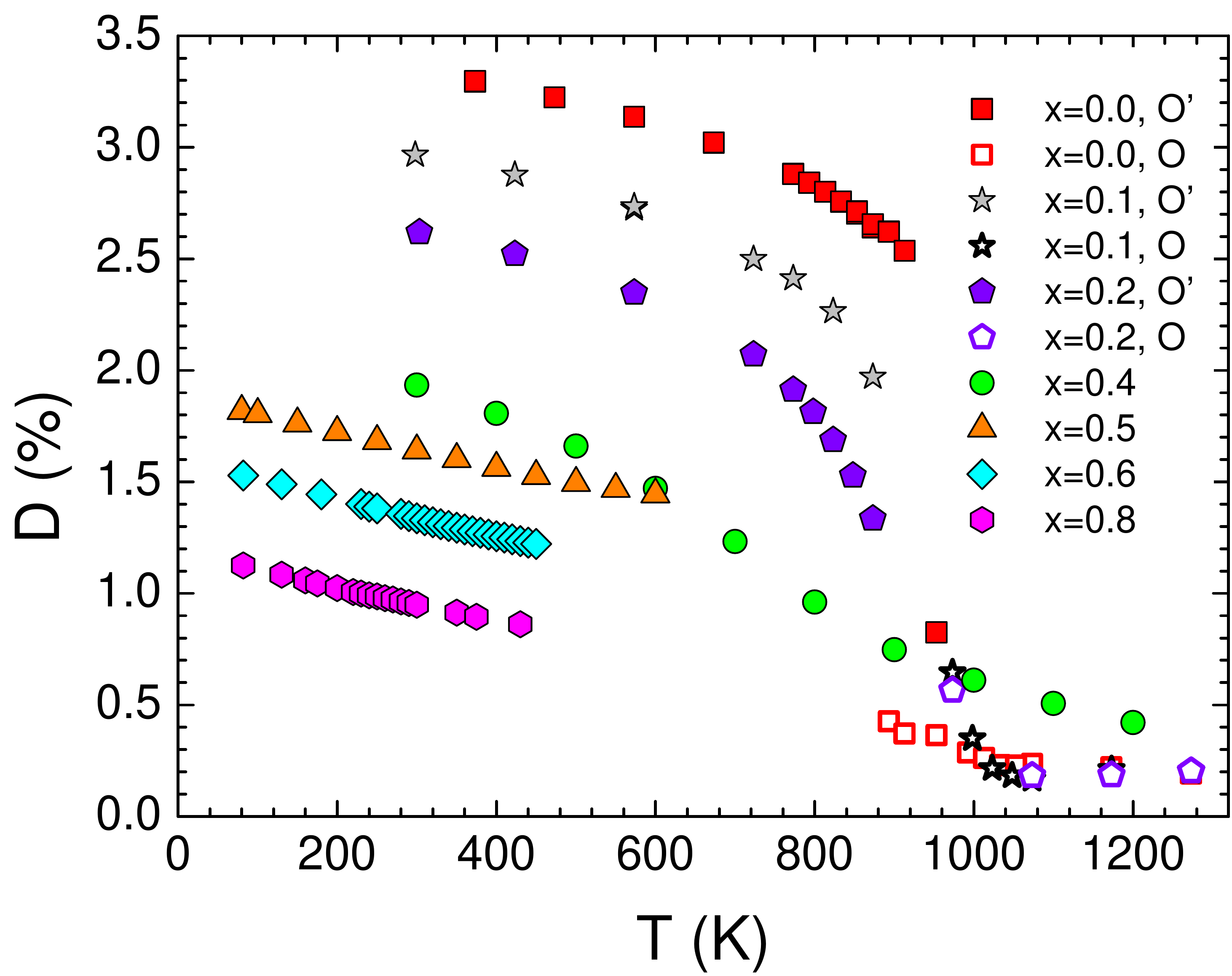}
\pdfcompresslevel=9
\caption{(Color online) The temperature dependence of the lattice distortion index $D$ for different iron concentrations $x$. \label{fig4}}
\end{center}
\end{figure}

One of the characteristics of O phase is $D$ parameter approaching zero value (Fig. \ref{fig4}). Such a structures were in older papers considered as quasi-cubic, or double-cubic systems, however, this notation is incorrect and the structure, as we have shown, has to be still considered as orthorhombic one.
 
The O' -- O structural phase transition is robust for the iron substitution $x \leq 0.2$ and disappears for $x \geq 0.4$. For concentrations $x \geq 0.4$ we have observed only O' phase at temperatures lower than 1200 K. For concentration range $0.4 \leq x \leq 0.6$ we have observed two different regimes for the pseudocubic parameters: $a_{pc} > c_{pc} > b_{pc}$ at lower temperatures and $a_{pc} > b_{pc} > c_{pc}$ at higher temperature (see Fig. \ref{fig2} and Supplementary online material, section 2). Both regimes are within the O' crystallographic phase and the switching temperature between these two regimes decreases with increasing of iron concentration.  Such an effect is understandable since one replaces the Mn$^{3+}$ ions which are JT active with Fe$^{3+}$ ions, which are not JT active. 
%
%
%
%
%
%
\begin{figure}[t]
\begin{center}
\includegraphics[angle=0,width=0.47\textwidth]{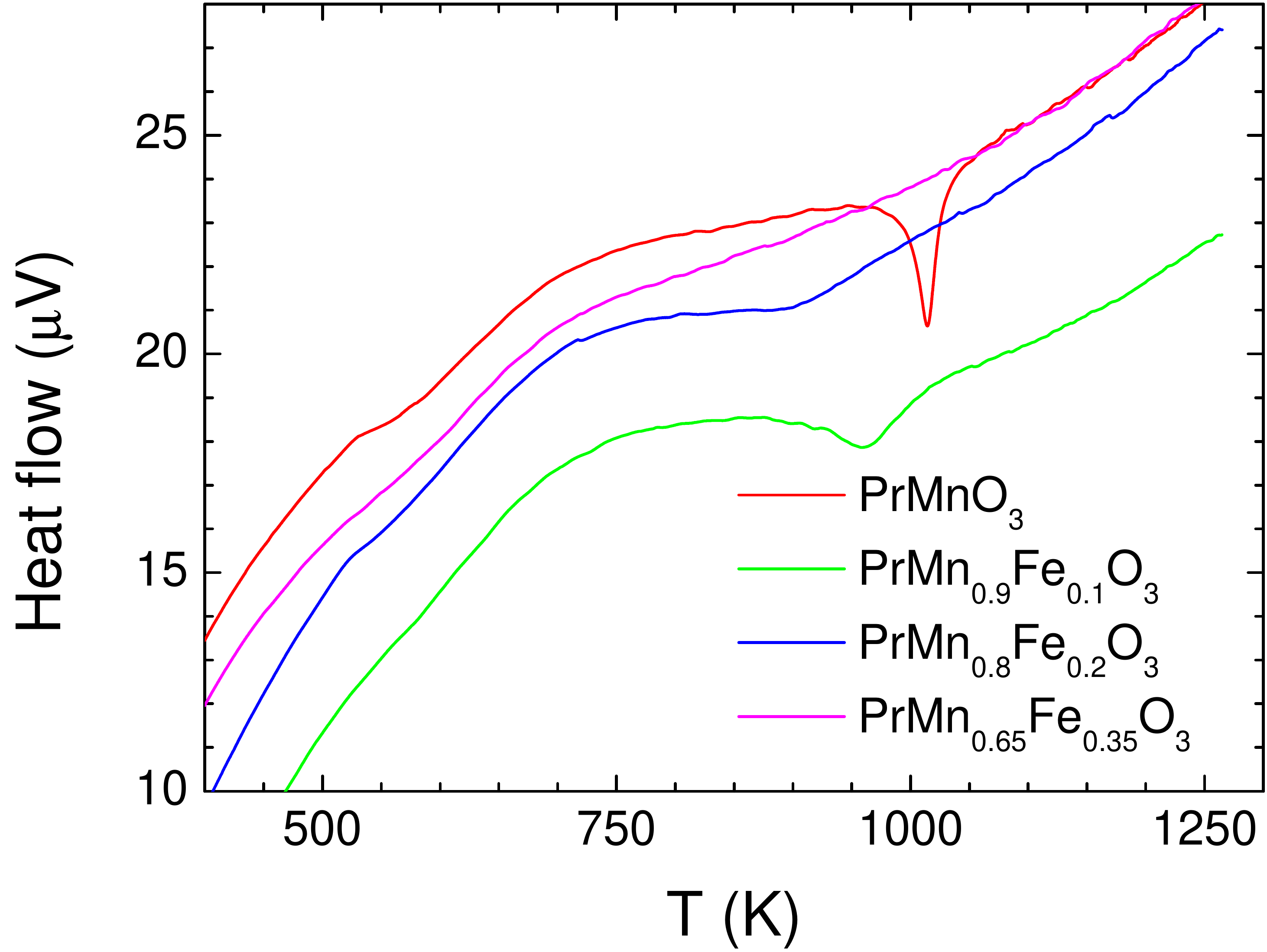}
\pdfcompresslevel=9
\caption{(Color online) DTA results for concentrations $0 \leq x \leq 0.35$. \label{fig5}}
\end{center}
\end{figure}
%
%

For all concentrations $D$ decreases with increasing temperature (Fig. \ref{fig4}) and $D$ decreases also for increasing of iron content. It means that increasing of iron concentration and temperature brings the crystal structure of PrMn$_{1-x}$Fe$_x$O$_3$ closer to the crystallographic structure of original perovskite.

XRPD measurements were performed with roughly 20 K step. Since this is quite high step in temperature sampling, we have determined the position of O' -- O structural phase transitions also by differential thermal analysis (DTA). DTA measurements show minima at 975(5) K; 928(10) K and 900(15) for $x$ = 0; 0.1 and 0.2, respectively but there is no anomaly for $x$ = 0.35 (Fig. \ref{fig5}). The minima are fingerprints of endothermic phase transitions and we define the transition as the onset of the minima. All these phase transitions coincide well with the O' -- O structural phase transitions as determined by XRPD experiments. 

Thermo gravimetry measurements (TG) revealed no anomaly in the studied temperature range. Such a result implies that the oxygen stoichiometry is constant in temperatures up to 1250~K and in air atmosphere.

Some more insights into the JT distortion could be done by examining directly the MnO$_6$ octahedrons in the studied compounds, however, in the Pnma orthorhombically distorted perovskite structure the oxygen atoms occupy $4c$ and $8d$ crystallographic positions which results to 5 adjustable parameters and the X-ray scattering cross section of oxygen is low. That is why our determination of oxygen fractional coordinates in the unit cell was too uncertain to present accurate results of (Mn,Fe)O$_6$ octahedrons metrics and Mn--O--Mn angles. For such a research the synchrotron or neutron experiment is highly desirable. However, the indirect study which was performed by comparing the lattice parameters shed at least some light to the problematic of the JT distortion in the PrMn$_{1-x}$Fe$_x$O$_3$ system.

%
%
\begin{table*}[t]
\begin{center}
\begin{tabular}{|cc|cccc|cc|c|}
\hline
&&\multicolumn{4}{c|}{ Fitting of phonons}&	\multicolumn{2}{c|}{Fitting of Schottky level}& \\
	
x&	fitting interval (K)&	$\theta_D$ (K)&	$\theta_{E_1}$ (K)&	$\theta_{E_2}$ (K)&	$\theta_{E_3}$ (K)&	$n_{\rm Sch}$&	$\Delta$  (K) &			$T_{\rm ord}$ (K)\\	
\hline
0&\begin{tabular}{c}2 - 50\\ 110 - 220\end{tabular}&184&216&503&690&0.75&17.6&95.2(3)\\
\hline
0.1&\begin{tabular}{c}2 - 50\\100 - 220\end{tabular}&170&229&498&647&0.75&19.4&82.3(3)\\
\hline
0.2&\begin{tabular}{c}0 - 35\\85 - 220\end{tabular}&164&239&477&633&0.74&20.8&62.5(5)\\
\hline
0.3&\begin{tabular}{c}0 - 30\\70 - 220\end{tabular}&162&245&466&571&0.73&21.5&38.4(5)\\
\hline
0.35&2 - 220&167&254&459&611&0.69&21.3&-\\
\hline
0.4&2 - 220&171&252&514&514&0.71&22&-\\
\hline
0.5&2 - 220&176&256&520&521&0.7&227&-\\
\hline
0.6&2 - 220&179&262	&525&525&0.66&21.4&-\\
\hline
0.8&2 - 230&188&259&545&545&0.66&21&-\\
\hline
\end{tabular}
\caption{The fitting parameters for the specific heat data. The regions where the lambda anomalies occur are omitted from the fitting. \label{table1}}
\end{center}
\end{table*}
%
%
%
%
\begin{figure}[t]
\begin{center}
\includegraphics[angle=0,width=0.47\textwidth]{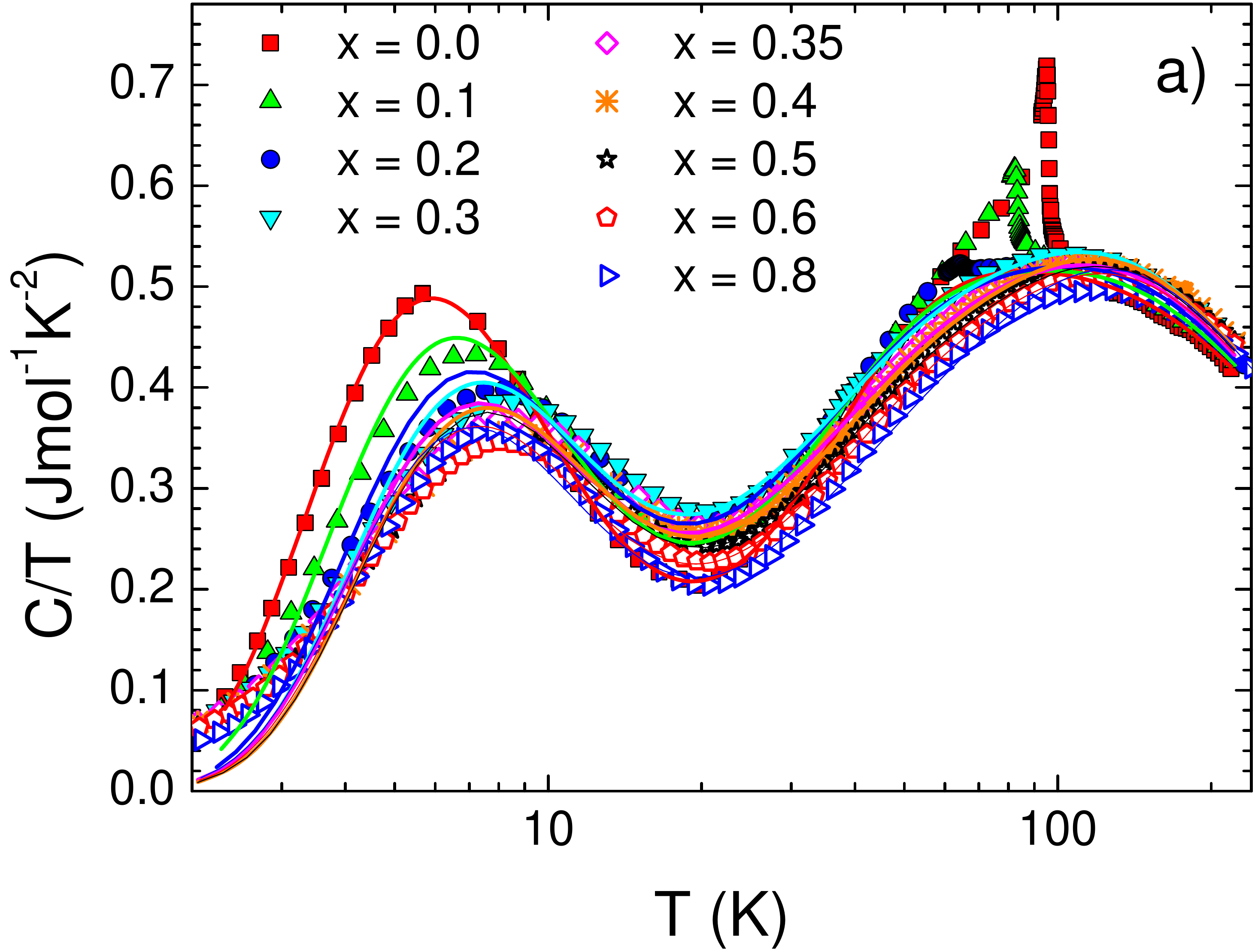}
\pdfcompresslevel=9
\includegraphics[angle=0,width=0.47\textwidth]{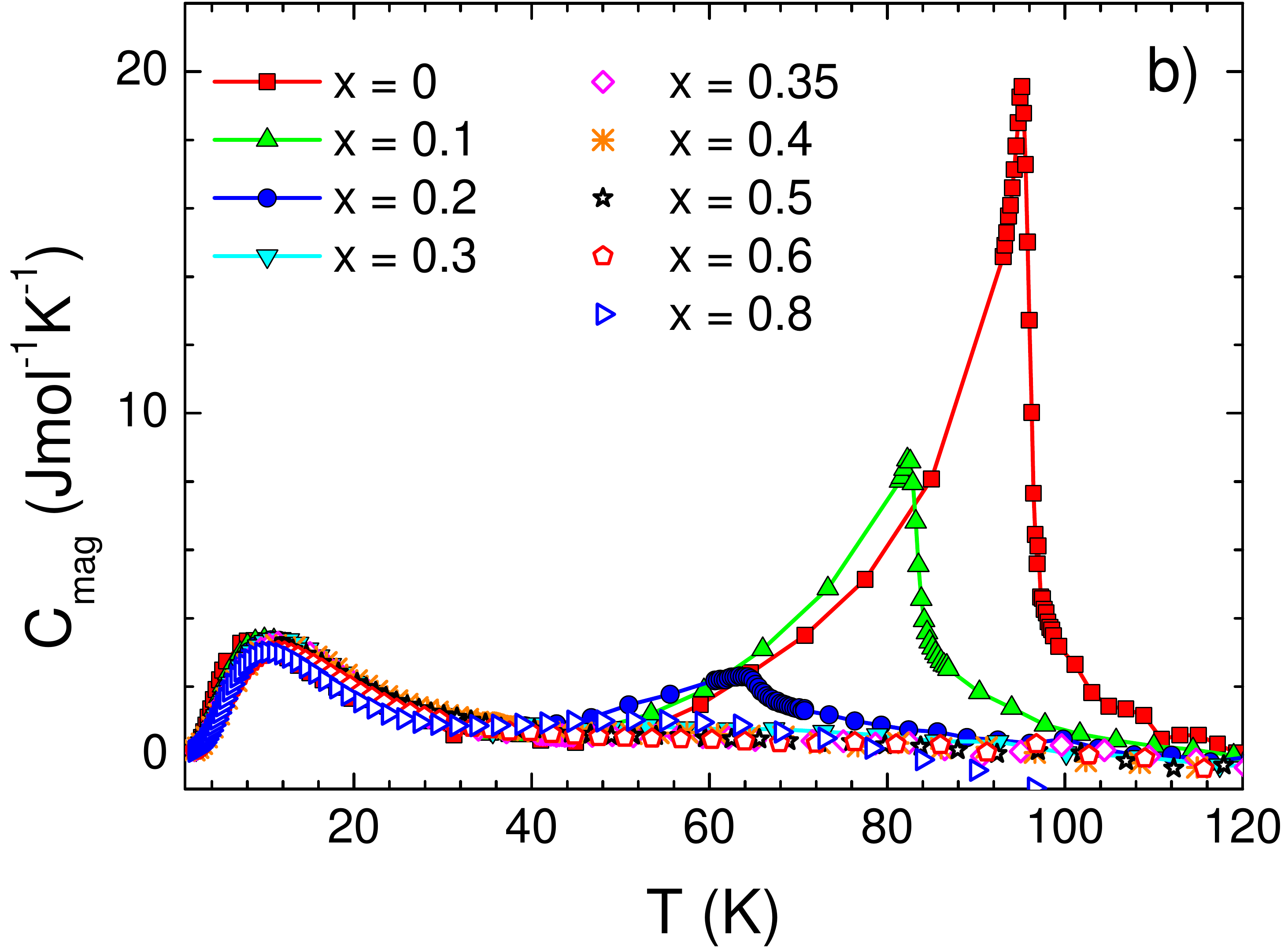}
\pdfcompresslevel=9
\caption{(Color online) a) the total specific heat of PrMn$_{1-x}$Fe$_x$O$_3$ compounds; lines represent the best fit of phonon contribution and Schottky levels as described in text b) the magnetic part of specific heat; lines are guides for the eye. \label{fig6}}
\end{center}
\end{figure}
%
%
\subsection{Heat capacity measurements}
Molar specific heat data in $C/T$ representation for compounds with $x \leq 0.3$ (Fig. \ref{fig6}a)) show phonon contribution and two distinct anomalies: the bump at around 10 K and the lambda-shape peak at higher temperatures. The specific heat data for concentrations $x > 0.3$ show only the phonon contribution, and bump around 10 K. The phonon contribution was modelled using one Debye term (characteristic temperature $\theta_{D_1}$) and three Einstein terms with multiplicity 4 for each term (characteristic temperatures $\theta_{E_1}$, $\theta_{E_2}$ and $\theta_{E_3}$, respectively). Following the approach of Hemberger et al. \cite{hemberger2004} who has studied the specific heat of PrMnO$_3$, we have treated the low temperature bump as 2-level Schottky contribution to the specific heat $(C_{\rm Sch})$:
\begin{equation}
C_{\rm Sch}=n_{\rm Sch} R\left(\frac{\Delta}{T}^2\right) \frac{\exp\left(\Delta/T\right)}{1+\exp\left(\Delta/T\right)^2 }
\label{CSch}
\end{equation}
where $R$ is the gas constant, $\Delta$ is the splitting of the Schottky levels, and $n_{\rm Sch}$ is the coefficient taking into account the number of Schottky centers as well the level of their degeneracy \cite{wahl2002}.  The results of the fit are presented in Table \ref{table1} and the match between the experimental data and fit is visualized in Fig. \ref{fig6}a. The results show that $\theta_D$ is the lowest around concentration $x$ = 0.3, while $\theta_{E_1}$ and $\theta_{E_2}$ increases and $\theta_{E_3}$ decreases with increasing concentration of iron. The shift of phonon characteristic temperatures can indicate changes of the geometry of (Mn,Fe)O$_6$ octahedrons, presumably due to removing of JT-distortion by the not JT active Fe$^{3+}$ ion. Schottky $\Delta$ increases with the iron substitution only moderately, while $n_{\rm Sch}$ moderately decreases, however, it is present in the whole concentration range. That is why we ascribe this effect to disordered Pr ions in the system. The magnetic contribution to specific heat (Cmag) as presented in Fig. \ref{fig6}b was obtained by the subtraction of the phonon contribution from the total specific heat. The typical features of $C_{\rm mag}(T)$ are lambda peaks. These lambda anomalies shift to lower temperatures with increasing concentration of iron and disappear for $x > 0.3$. We ascribe these anomalies to disorder-to-order magnetic phase transition and we refer the transition temperature to the maximum of the lambda peak. The results of the data processing suggest that the ordering temperature decreases with increasing content of iron (see Table \ref{table1}) from 95.2(3) K for PrMnO$_3$ to 38.4(5) K for $x$ = 0.3.

\begin{figure}[t]
\begin{center}
\includegraphics[angle=0,width=0.47\textwidth]{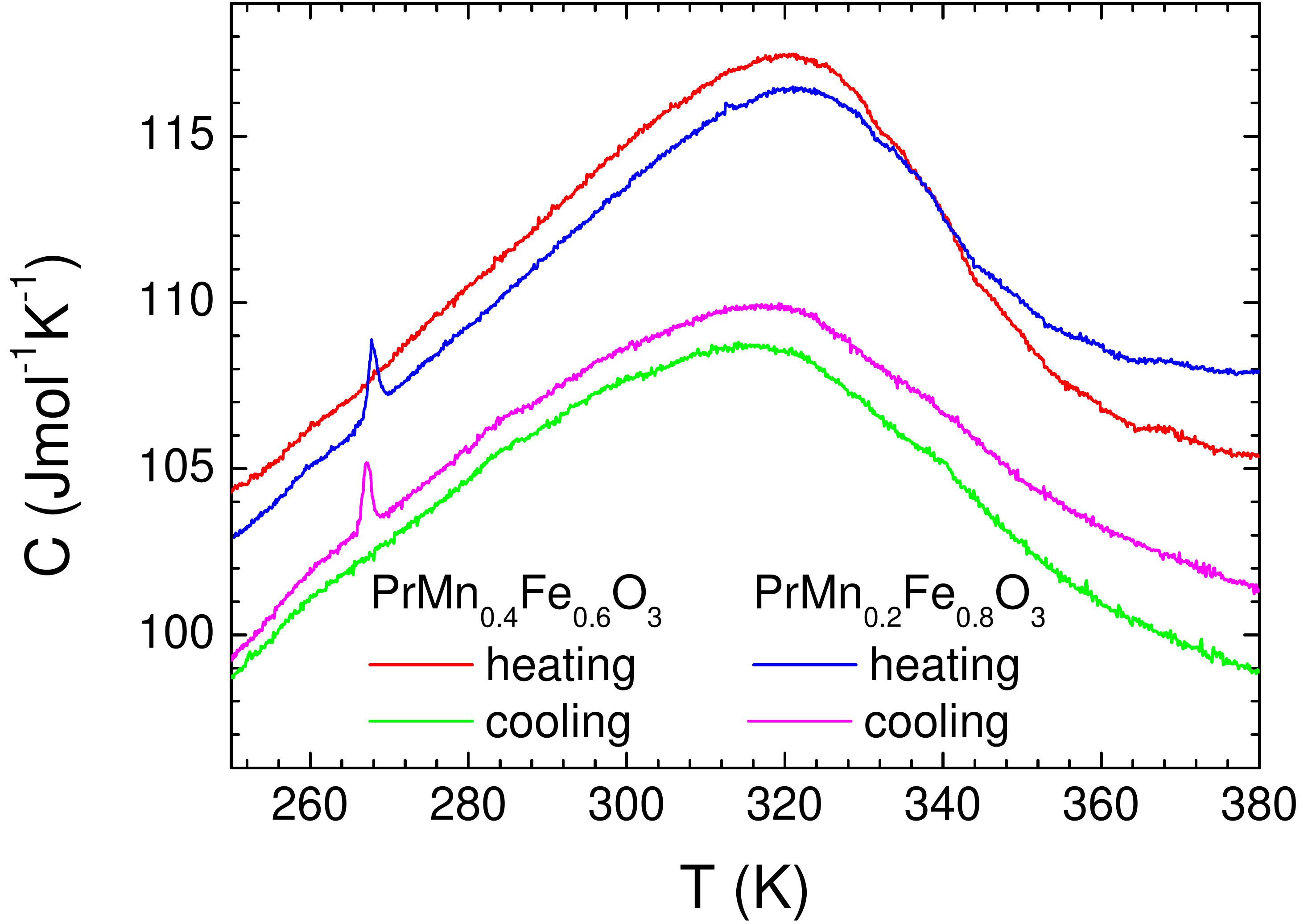}
\pdfcompresslevel=9
\caption{(Color online) Specific heat for $x$ = 0.6 and 0.8 as obtained by DSC method. \label{fig7}}
\end{center}
\end{figure}

It has been reported that PrFeO$_3$ orders magnetically well above the room temperatures \cite{pinto1972}. That is why one can expect that some magnetic phase transitions in PrMn$_{1-x}$Fe$_x$O$_3$ system exist at room temperatures, or higher. Relaxation method for determining the heat capacity presented in previous paragraphs is not very suitable method for temperatures higher than approximately 250 K, since the radiation losses of heat can obscure the result. That is why we have investigated the samples with concentration $x \geq 0.6$ also by differential scanning calorimetry (DSC) (Fig. \ref{fig7}). DSC signal from PrMn$_{0.2}$Fe$_{0.8}$O$_3$ shows peak on both, cooling and heating cycle. These peaks are slightly shifted, but this shift can be understood as the side effect of the experimental method (heating and cooling rate was 10 K/min). Nevertheless these peaks unambiguously identify that there is phase transition. We have determined the temperature of this phase transition as the arithmetic average of the position of the peak maximum in heating and cooling cycle and this transition takes place at 267.4(3) K. Our measurements show that in the case of PrMn$_{0.4}$Fe$_{0.6}$O$_3$ no anomalies in studied temperature interval 250 -- 390 K are present.
%
%

%
%
%
%
\subsection{Magnetic measurements} 
%
%
\begin{figure*}[t]
\begin{center}
\includegraphics[angle=0,width=0.95\textwidth]{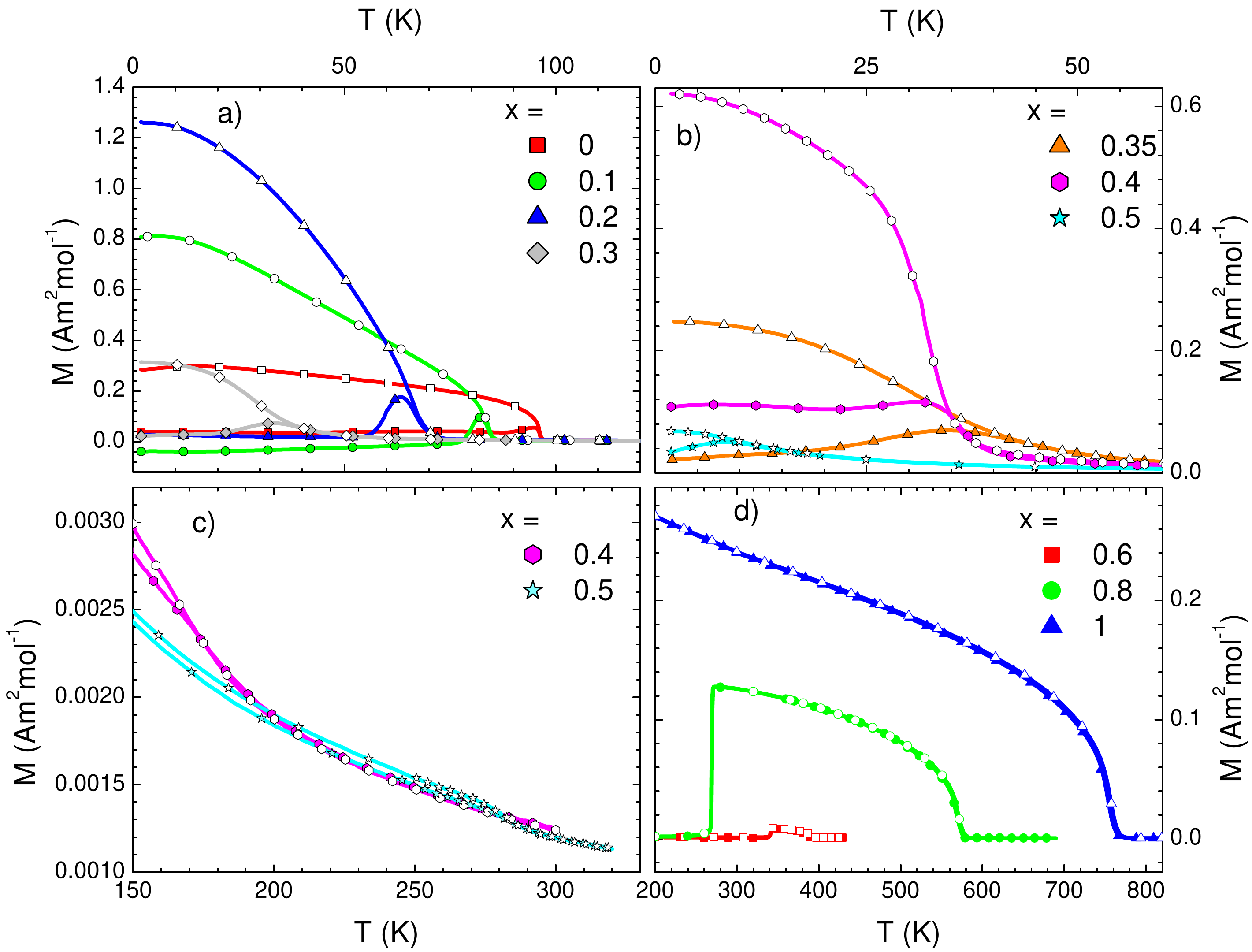}
\pdfcompresslevel=9
\caption{(Color online) The ZFC (full symbols) and FC (opened symbols) curves measured in applied magnetic field $\mu_0$H = 0.01 T. In the case of concentrations $x \leq 0.5$ [panels a), b, and c)] the data were obtained on powder samples, in case of concentrations with $0.5 \leq x$ [panel d)] the presented data were obtained on single crystals with magnetic field applied along the $b$-axis. The data obtained along the remaining two crystallographic axes for concentrations $x = $0.5; 0.6; 0.8 and 1 can be found in Supplementary online material, section 3. \label{fig8}}
\end{center}
\end{figure*}
%
%

\begin{figure}[t]
\begin{center}
\includegraphics[angle=0,width=0.47\textwidth]{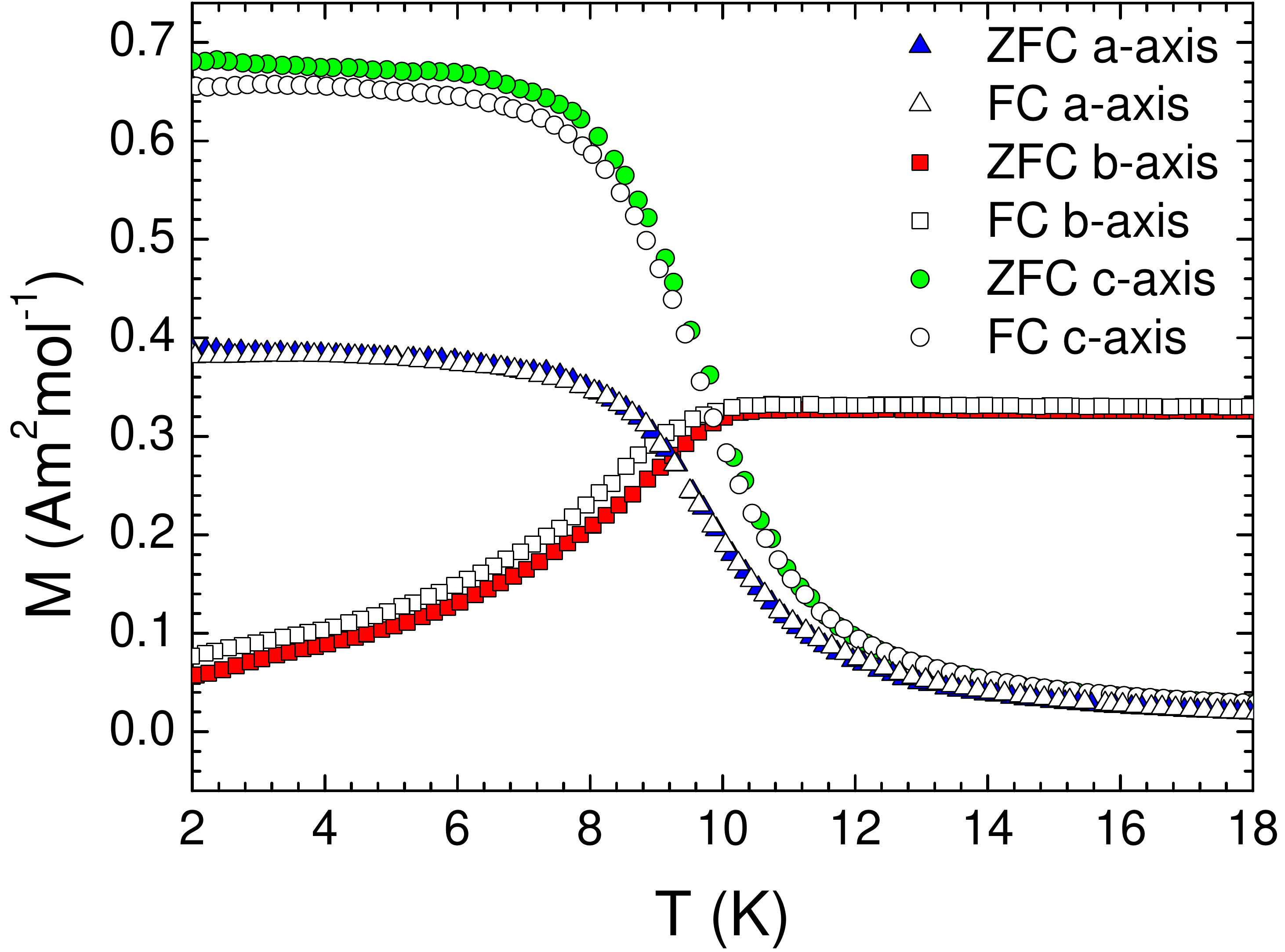}
\pdfcompresslevel=9
\caption{(Color online) The detail of the ZFC and FC magnetization curves of PrFeO$_3$ for measurements in applied field with induction of 0.01 T applied along all three main crystallographic axes. \label{fig9}}
\end{center}
\end{figure}

Zero-field-cooled (ZFC) and field-cooled (FC) magnetization measurements (Fig. \ref{fig8}) revealed that ZFC bifurcates from FC curve for all concentrations with $x \leq 0.5$. For concentrations $x > 0.5$ the ZFC and FC curves do not bifurcate, however the shape of the curves unambiguously indicate the paramagnetic-to-magnetically ordered phase transition takes place at temperatures higher than room temperature. The magnetic ordering transition we define as the minimum on (dM(T)/dT) curve and the transition takes place at 95.9(3) K; 84.4(3) K; 66.9(3) K; 39.0(7) K; 42.9(4); 170(1) K and 283(1) K  for $x$ = 0; 0.1; 0.2; 0.3; 0.35; 0.4 and 0.5, which is slightly lower than the bifurcation point. In the case of concentration $x$ = 0.5 we have obtained the identical results for single crystalline and powder samples. In the case of the single crystalline data we have obtained slightly different transition temperatures for measurements with  magnetic field oriented along different crystallographic directions (see Supplementary data). In order to compare data from polycrystalline and single crystalline samples we determine the magnetic phase transition on single crystals as the arithmetic average of relevant ordering temperatures, which amounts 280(1) K; 391(5) K; 570(5) K and 738(5) K for $x$ = 0.5; 0.6; 0.8 and 1, respectively. The additional anomalies characterized by the increase of magnetization, were found on M(T) curves for $x$ = 0.4 and 0.5 at $T_1$ = 35.5(2) K and 12.0(2) K. Large drop of magnetization in temperature interval 338 -- 355 K and sharp drop of magnetization at $T_2$ = 269(1) K for $x$ = 0. 6 and for $x$ = 0.8, respectively, are another typical magnetic anomalies in PrMn$_{1-x}$Fe$_x$O$_3$. The latest anomaly matches well with the anomaly observed by DSC. In the case of PrFeO$_3$ we have observed the drop of the magnetization at 9.5 K, in the case of measurements with magnetic field applied along $b$-axis, but increase of the magnetization at the same temperature when magnetic field was applied along $a$- and $c$-axis (Fig. \ref{fig9}). Such a behavior implies that the rotation of the magnetic spin away from $b$-axis takes place at this transition.
%
%
\begin{table*}[t]
\begin{center}
\begin{tabular}{ccccc}
\hline
x&	fiting interval (K)& $\mu_{\rm eff} \mu_B$&	$\theta_{\rm P}$ (K)&	$\chi_0$ (m$^3$mol$^{-1}$)\\
\hline
0&	120 -- 300&	7.07(2)&	-17.8(5)&	-3.04(8)$\times10^{-8}$\\
0.1&	120 -- 300&	6.07(1)&	10.6(2)&	-8.4(1)$\times10^{-10}$\\
0.2&	120 -- 300&	5.50(1)&	25.9(2)&	1.36(3)$\times10^{-8}$\\
0.3&	120 -- 300&	4.61(1)&	42.8(6)&	3.41(9)$\times10^{-8}$\\
0.35&	120 -- 380&	4.77(1)&	45.0(1)&	4.1(1)$\times10^{-8}$\\
\hline
\end{tabular}
\caption{The results of fitting the susceptibility data due to modified Curie-Weiss law. \label{table2}}
\end{center}
\end{table*}
%
%
%
%
%
%
%
%
\begin{figure}[t]
\begin{center}
\includegraphics[angle=0,width=0.47\textwidth]{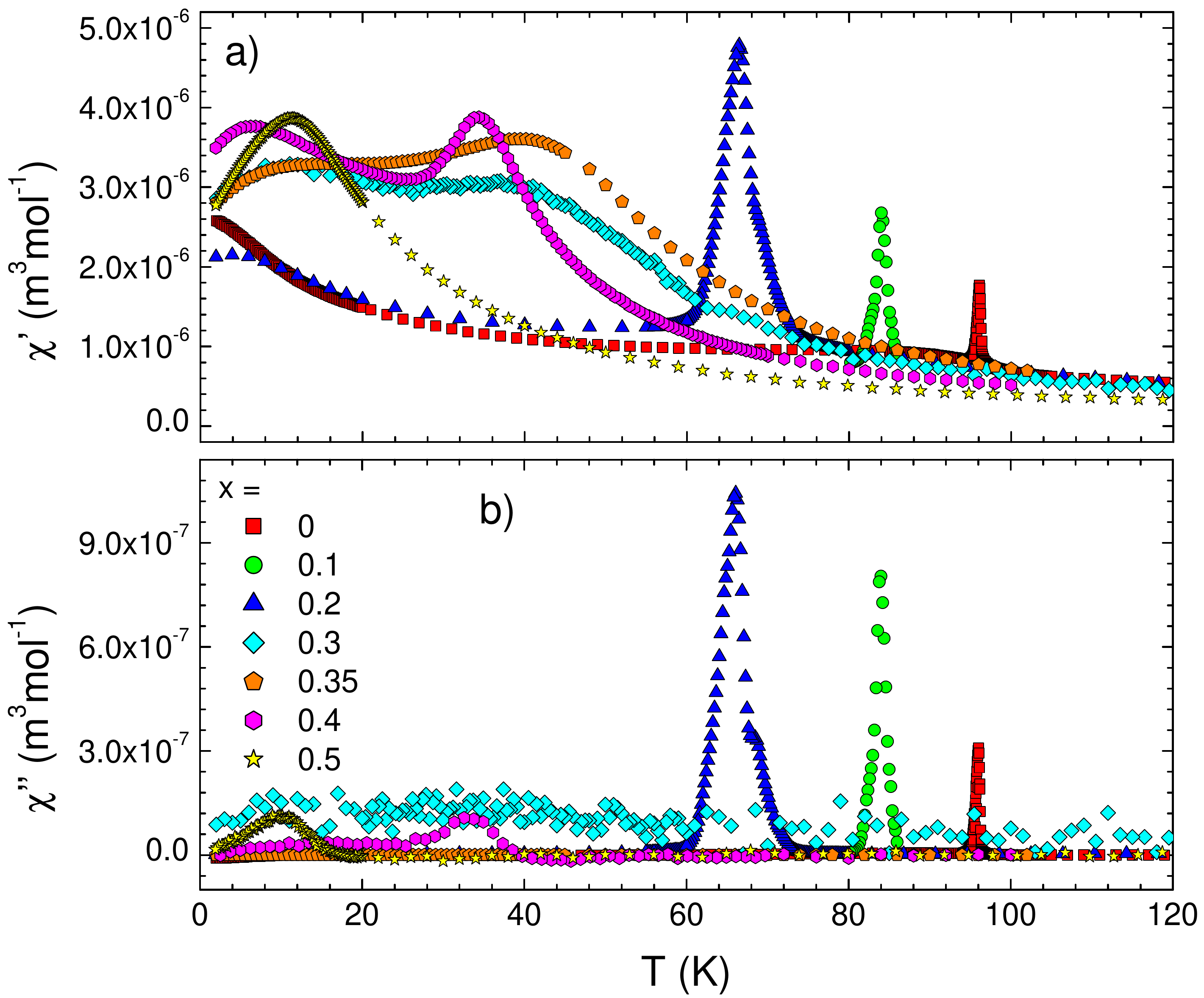}
\pdfcompresslevel=9
\caption{(Color online) a) the real part and b) the imaginary part of AC susceptibility for measurements at frequency of 1 kHz and with driving field 6.5 Oe (x = 0; 0.1 and 0.2) or 1 Oe (x = 0.3; 0.35; 0.4 and 0.5). \label{fig10}}
\end{center}
\end{figure}
%
%

\begin{figure}[t]
\begin{center}
\includegraphics[angle=0,width=0.47\textwidth]{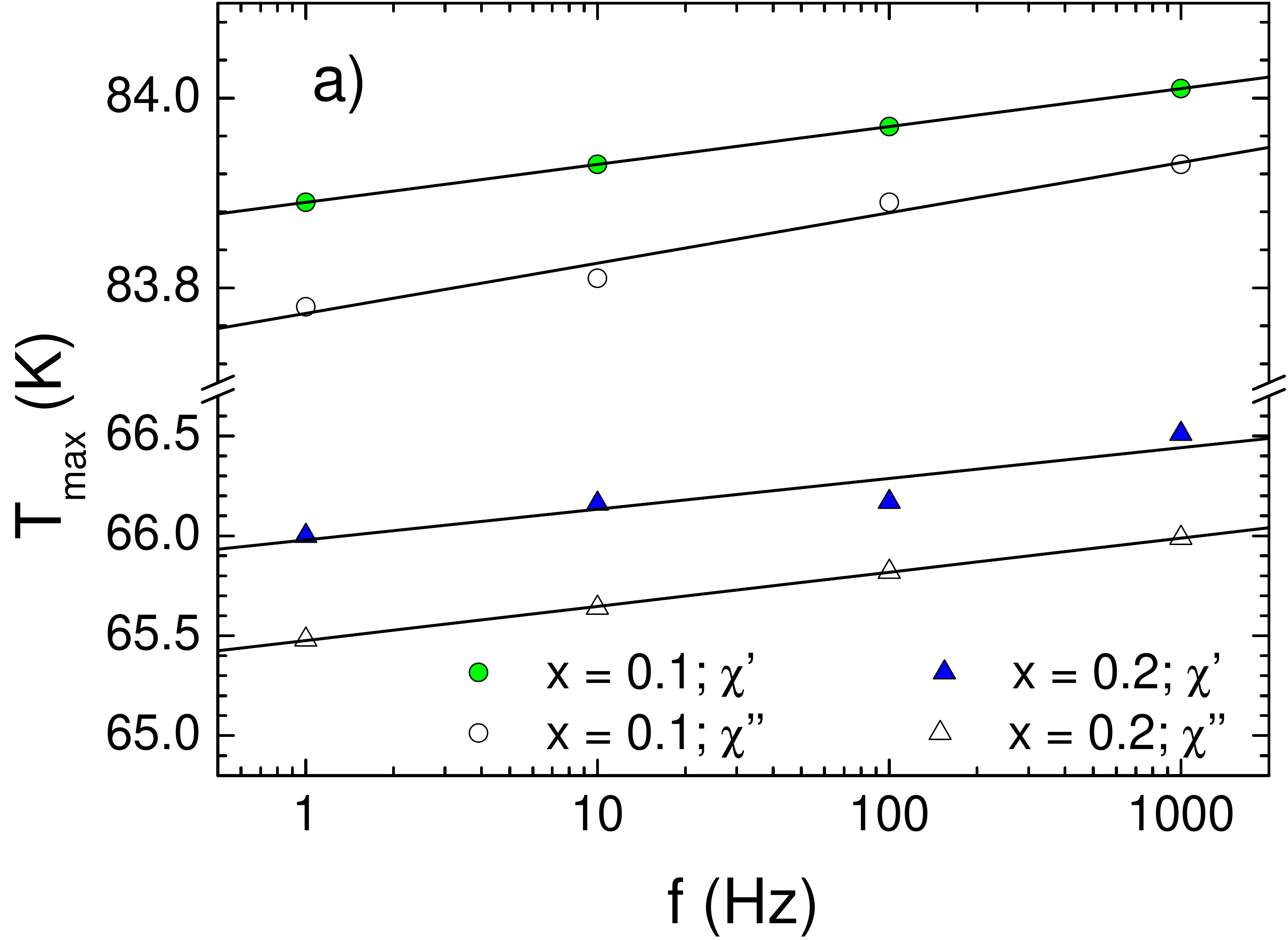}
\pdfcompresslevel=9 \\
\includegraphics[angle=0,width=0.47\textwidth]{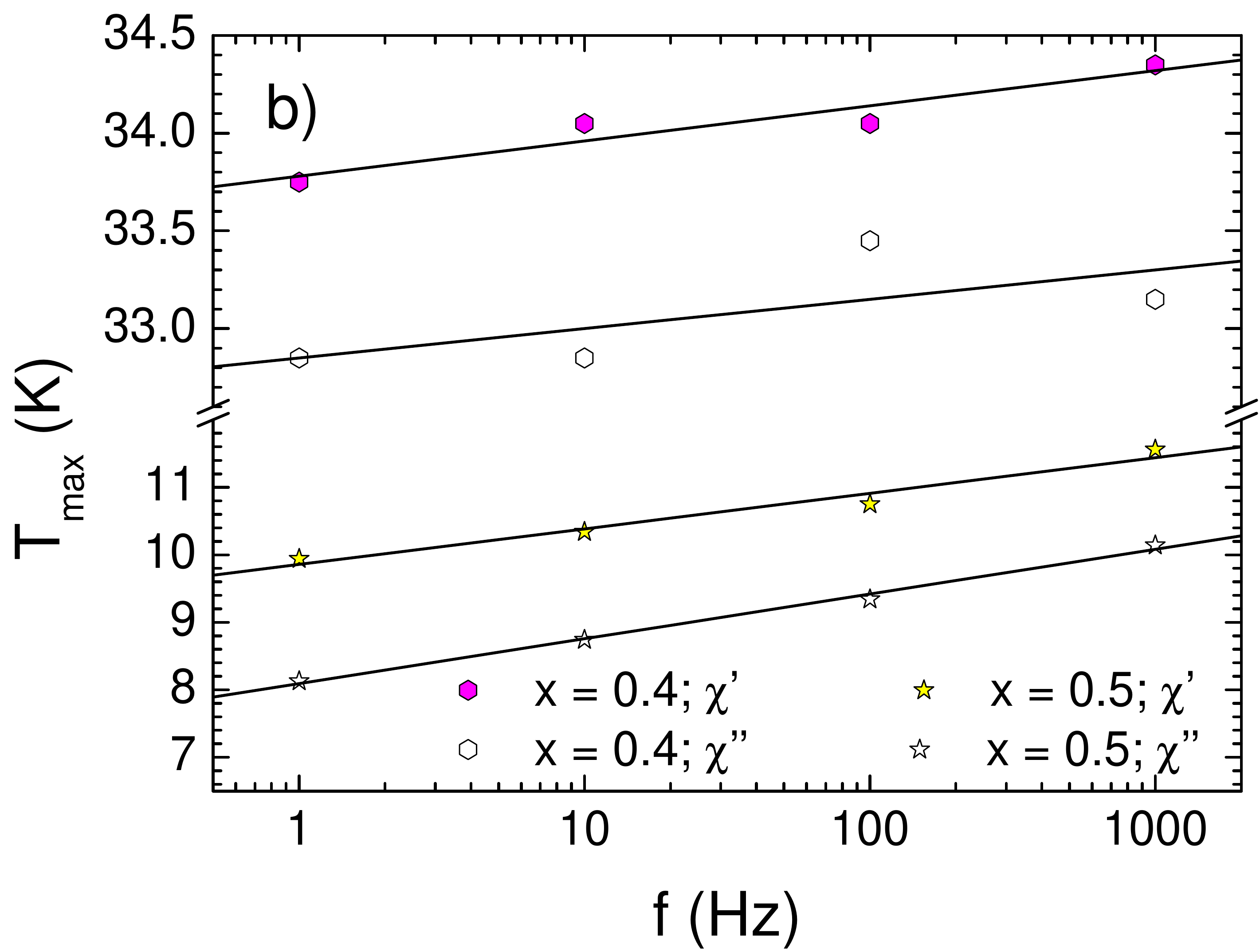}
\pdfcompresslevel=9
\caption{(Color online) The frequency dependence of anomalies in AC susceptibility for a) $x$ = 0.1 and 0.2 b) $x$ = 0.4 and 0.5. The lines represent the linear fit in log(f) scale as described in text. \label{fig11}}
\end{center}
\end{figure}

Measurements of AC susceptibility on PrMnO$_3$ revealed sharp peaks at about 96 K in both, the real part of AC susceptibility $\chi'_{\rm AC}$  and the imaginary part of AC susceptibility $\chi''_{\rm AC}$ (Fig. \ref{fig10}). These peaks are frequency independent in the frequency range 1 Hz $\leq$ f $\leq$ 1 kHz. The maxima increase linearly in log(f) scale, become broader and shift to 84 K and 66 K with increasing iron content from $x$ =  0.1 to $x$ =  0.2 (Fig. \ref{fig11}), respectively. The broad bumps only in $\chi'_{\rm AC}$ at about 35 K and 40 K are typical features of samples with $x$ = 0.3 and 0.35. These bumps change the position slightly with increasing frequency. Another quite sharp maxima, which are again present in both $\chi'_{\rm AC}$ and $\chi''_{\rm AC}$, appear at about 34 K and 11.5 K for $x$ = 0.4 and 0.5, respectively.
  
Magnetic isotherms, which were measured at 2 K, exhibit hysteretic behavior for the whole series (Fig.\ref{fig12} and Supplementary online material, section 4). The hysteresis loops on powder samples for $x \leq 0.3$ exhibit kink in the vicinity of remnant magnetization (Fig. \ref{fig12}a). Such a behavior was previously observed in La$_{1-x}$Ca$_x$MnO$_3$ \cite{wollan1955} and NdMn$_{1-x}$Fe$_x$O$_3$ powder samples \cite{mihalik2013}. This shape of hysteresis loops indicates that the magnetic structure is complex and the interplay between ferromagnetic and antiferromagnetic interactions can be strong. Our interpretation is totally in agreement with the proposed magnetic structure C$_{\rm x}$F$_{\rm y}$ of Mn ions as proposed by Baran et al. \cite{baran2013}. The hysteresis loops for $0.3 < x < 0.5$ (Fig. \ref{fig12}b) are very similar with hysteresis loops from the previous region of concentration, but they do not content any kink in the vicinity of remnant magnetization. The ferromagnetic component is present and the loops look like superposition of antiferromagnetic and ferromagnetic contribution. The hysteresis loops for $0.5 < x \leq 1$ in the case of single crystals (Fig. \ref{fig12}c) are much sharper, which is the consequence of the quality of the sample and the coercive force is lower than in the case of powder samples $(0 < x < 0.5)$. Nevertheless, hysteresis persists when measured along all three main crystallographic directions (Fig. \ref{fig12}c and the Supplementary online material, section 4) which suggest that ferromagnetic interactions are also present at relevant temperatures. The coercive forces $\mu_0$H = 0.09 T; 0.11 T; 0.12 T for main crystallographic axes ($a$-; $b$- and $c$-axis) of single crystal and $\mu_0$H = 0.07~T determined from and the powdered sample with $x$ = 0.5 at 2 K are very similar but still there are some differences among them indicating magnetocrystalline anisotropy of the material (see Supplementary online material). The hysteresis loops around $T_1$ (Fig. \ref{fig12}d and Supplementary online material) exhibit two distinct anomalies: the hysteresis behavior around zero magnetic field with very low coercive force and metamagnetic-like step in higher magnetic fields. This implies onset of antiferromagnetic interactions at $T_1$. Also the hysteresis loops in temperature range $T_1 < T < T_{\rm ord}$ suggest the presence of ferromagnetic interactions which co-exist with mainly antiferromagnetic interactions. 

The inverse magnetic susceptibility at temperatures higher than 120 K is almost linear for compounds with $x \leq 0.35$ (Supplementary online material). That is why we have modelled the susceptibility for these concentrations with the modified Curie-Weiss law: 
\begin{equation}
\chi = \chi_0+\frac{C}{T-\theta_{\rm P}}
\label{CW}
\end{equation}
where $\chi_0$ is the temperature independent contribution to the measured susceptibility (presumably contribution from the sample holder and glue); $\theta_{\rm P}$ is the characteristic temperature and constant $C$ scales with the effective magnetic moment $\mu_{\rm eff}$. The fitted parameters (Table \ref{table2}) revealed that effective magnetic moment decreases linearly with increasing of iron concentration with rate -0.43(3) $\mu_{\rm B}$ per 10 \% of iron substitution, but in the same time the characteristic temperature increases with rate 18 K per 10 \% of iron substitution (Table \ref{table2}). On the other hand, Fe$^{3+}$ ion in the high spin state has one more unpaired electron if comparing with Mn$^{3+}$ ion. This implies that effective moment should increase with iron substitution. Since the opposite effect was observed, it is possible that Curie-Weiss law is not fulfilled for the PrMn$_{1-x}$Fe$_x$O$_3$ system. The reason for this might be also the high magnetocrystalline anisotropy in the system. The high magnetocrystalline anisotropy was proven for concentrations $0.5 \leq x \leq 1$.
%
%
%
%

\section{Summary and Discussion}
Experimental results as presented in previous sections allowed us the construction of the structural and magnetic phase diagram (Fig. \ref{fig13}). We have observed O' -- O structural phase transition at $T_{\rm JT}$ for concentration up to $x$ = 0.2. The degeneracy of the $e_g$ orbital is removed by a cooperative Jahn-Teller distortion below $T_{\rm JT}$. This distortion results in a long-range ordering of the occupied $3d_{3x^2-r^2}$ and $3d_{3y^2-r^2}$ orbitals. Such a transition occurs for instance in PrMnO$_3$ compound at 1030 K \cite{zhou2003}. The transition from an orbital-disordered, via a short-range ordered regime, to a long-range orbital-ordered state takes place in the temperature range $T^* < T < T_{\rm JT}$. This intermediate temperature range is narrowing with increasing concentration of iron and one can extrapolate that $T^* \sim T_{\rm JT}$ for $x$ slightly higher than 0.2. We did not observed O' -- O structural phase transition for samples with higher concentrations than $x$ = 0.2. The O' phase exists in whole temperature range up to 1200 K for sample with $x \geq 0.4$. In addition, the change of relation $a_{pc} > c_{pc} > b_{pc}$ to $a_{pc} > b_{pc} > c_{pc}$ was observed up to $x$ = 0.6, which indicates that JT distortion is still present in this concentration range and can contribute to the distortion of (Mn,Fe)O$_6$. Temperature evolution of border between region JT, where Jahn-Teller distortion contributes to orthorhombic distortion of (Mn,Fe)O$_6$ octahedrons and $RE$, where the orthorhombic distortion is given dominantly by tilting and rotation of (Mn,Fe)O$_6$ octahedrons is plotted in Fig.\ref{fig13} for concentrations $0.4 \leq x \leq 0.6$.
%
%
\begin{figure*}[t]
\begin{center}
\begin{subfigure}{0.45\textwidth}
\includegraphics[angle=0,width=0.95\textwidth]{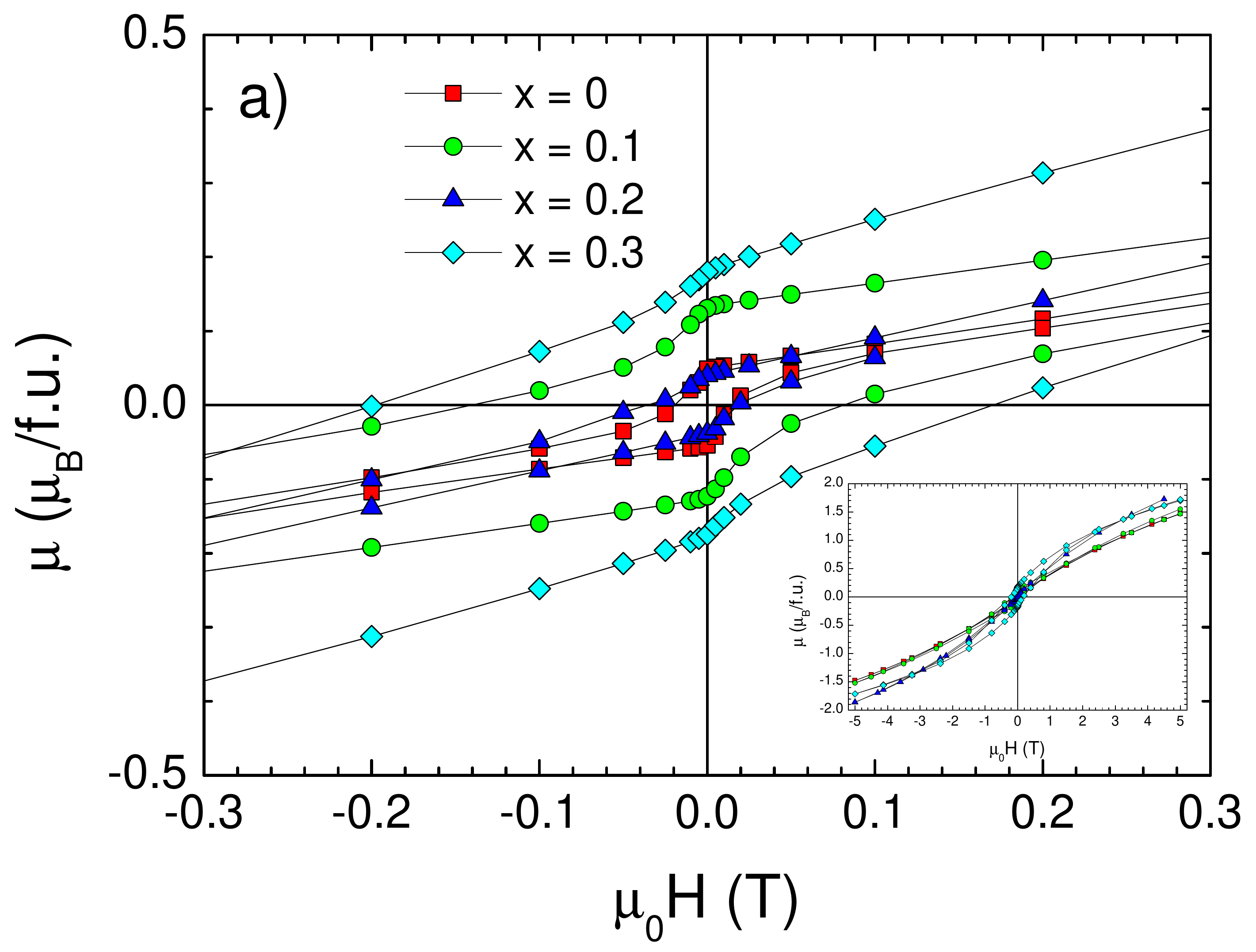}
\pdfcompresslevel=9
\end{subfigure} \hspace{0.05\textwidth}
\begin{subfigure}{0.45\textwidth}
\includegraphics[angle=0,width=0.95\textwidth]{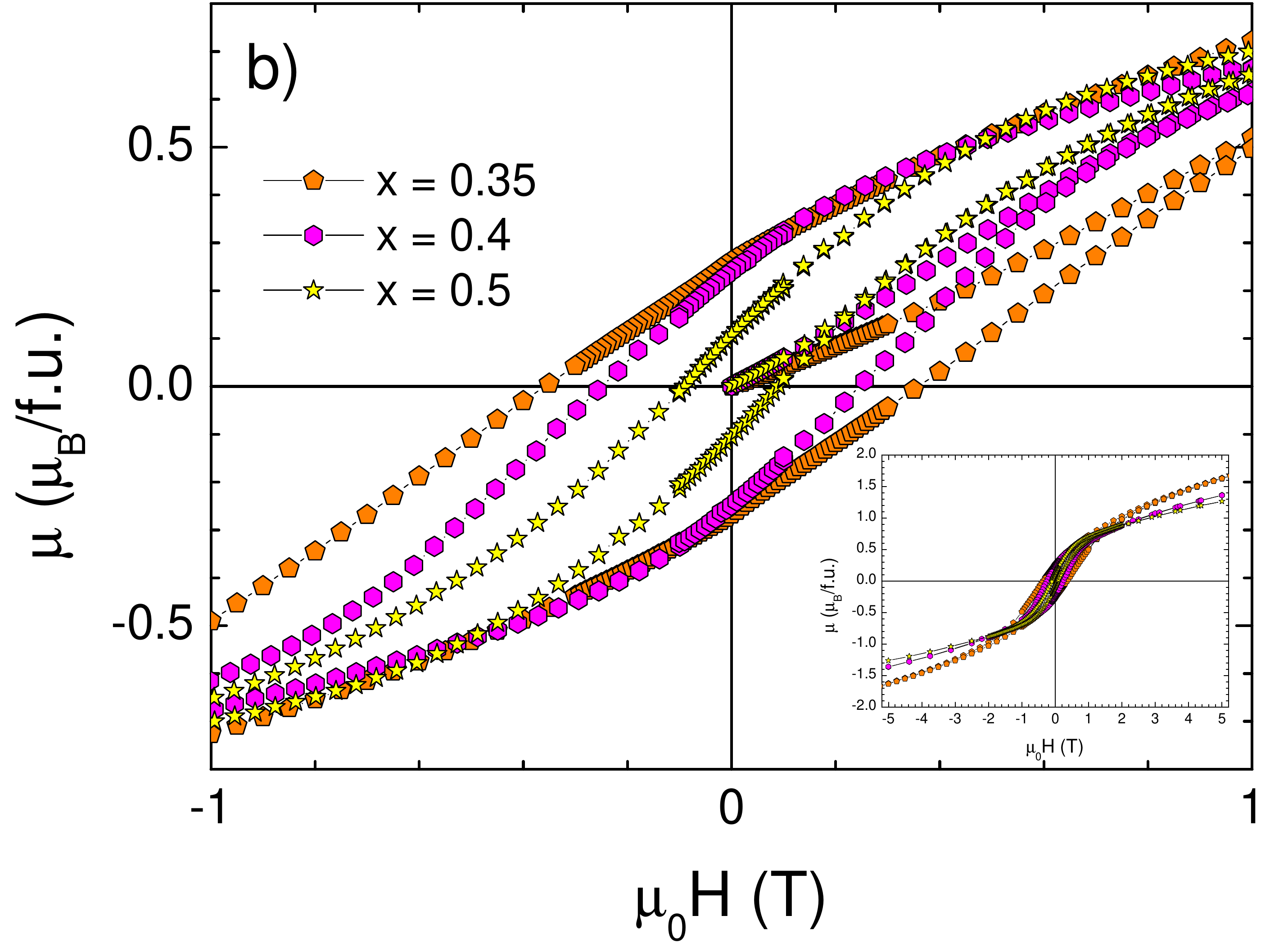}
\pdfcompresslevel=9
\end{subfigure} \\
\begin{subfigure}{0.45\textwidth}
\includegraphics[angle=0,width=0.95\textwidth]{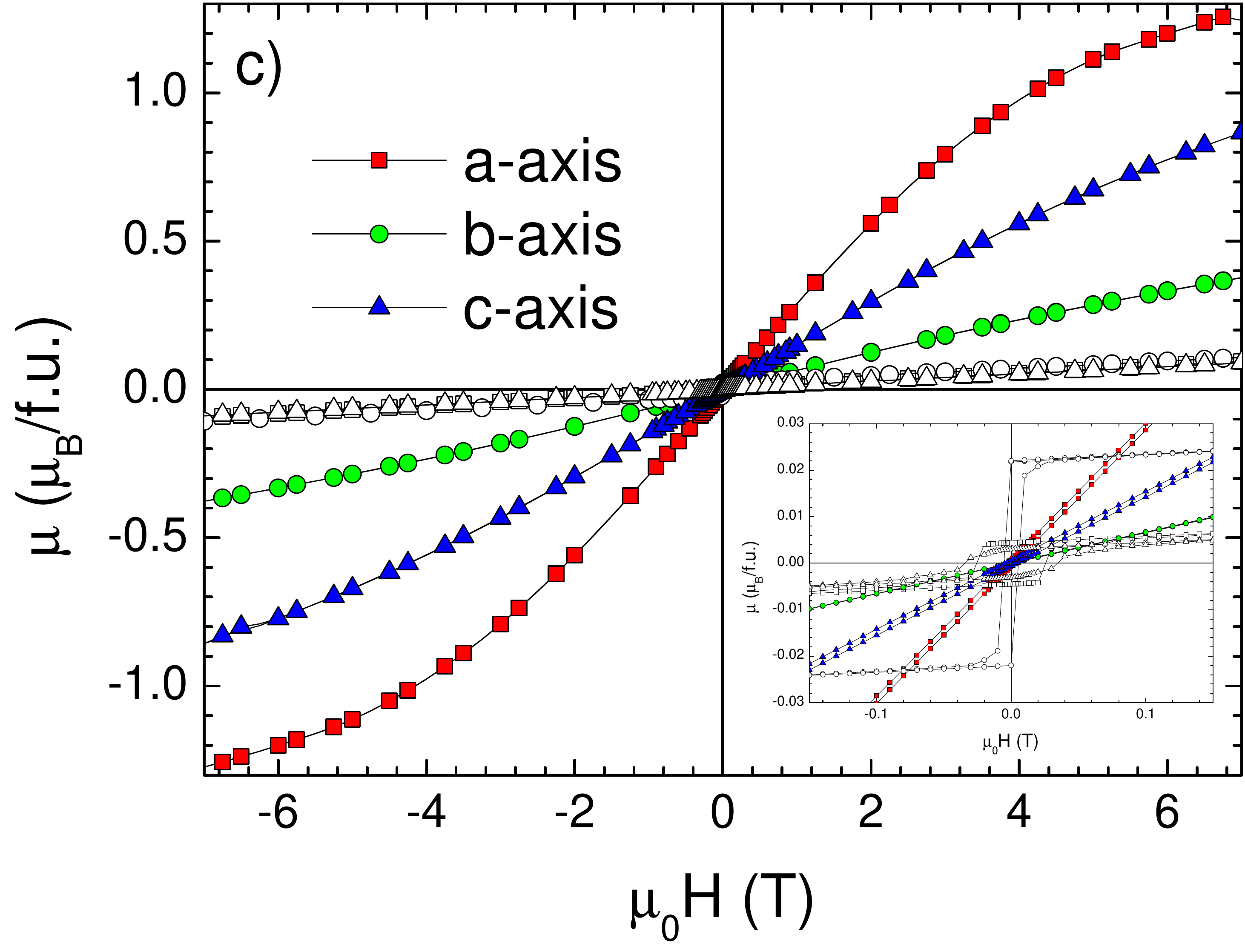}
\pdfcompresslevel=9
\end{subfigure} \hspace{0.05\textwidth}
\begin{subfigure}{0.45\textwidth}
\includegraphics[angle=0,width=0.95\textwidth]{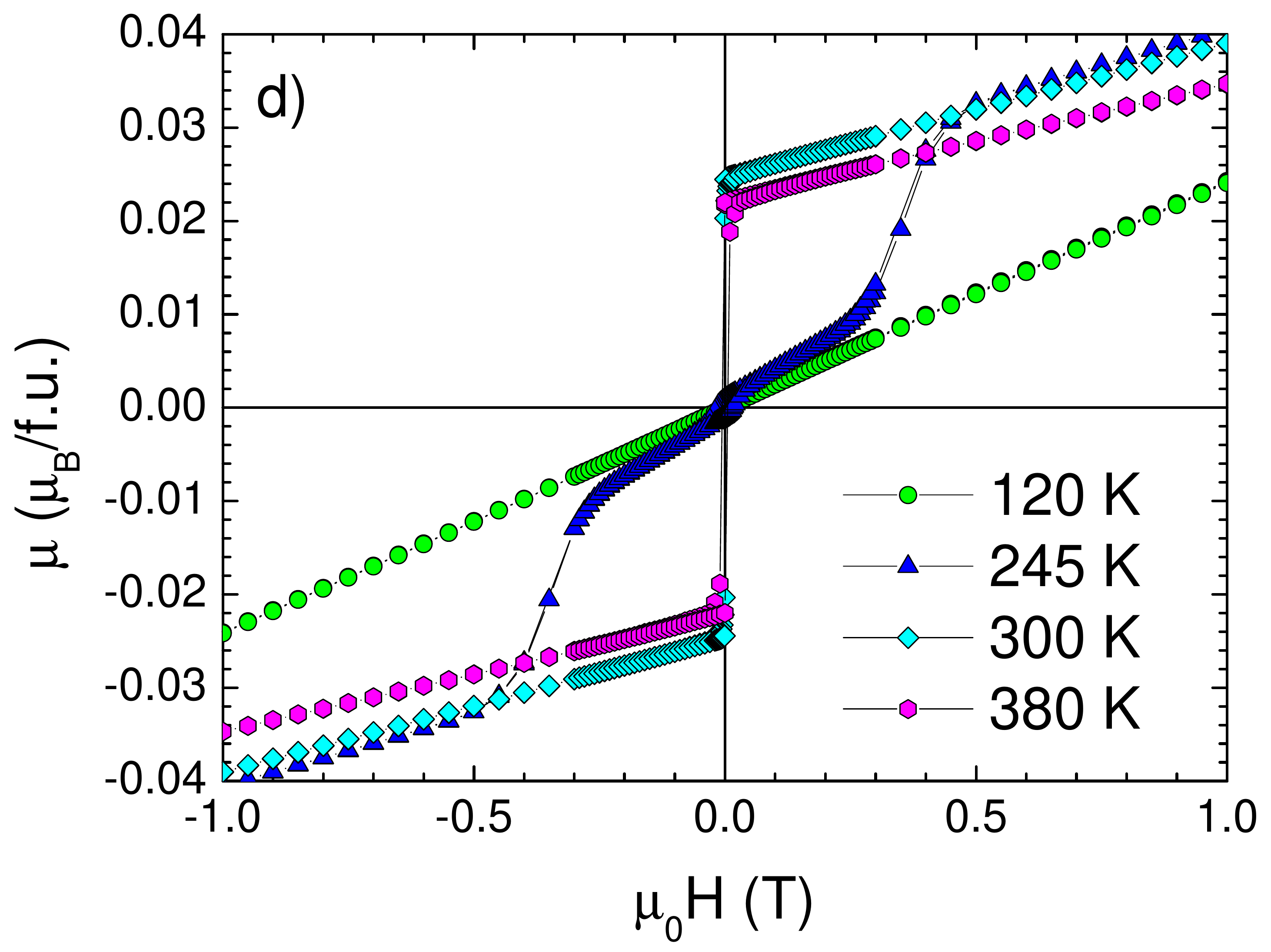}
\pdfcompresslevel=9
\end{subfigure}
\caption{(Color online) Hysteresis loops a) for $x \leq 0.3$ measured at T = 2 K b) for $0.35 \leq x \leq 0.5$ measured at $T$ = 2 K c) for $x$ = 0.8 measured at $T$ = 2 K (full symbols) and $T$ = 380 K (opened symbols) d) for $x$ = 0.8 and magnetic field applied along b-axis. a) and b) : measurements performed on powders; c) and d): measurements performed on single crystals. Data for compositions $x = $ 0.5; 0.6 and 1 are presented in the Supplementary online material, section 4. \label{fig12}}
\end{center}
\end{figure*}
%
%

Magnetic phase diagram is complex (Fig. \ref{fig13}). The paramagnetic to magnetic transition of both parent compounds PrMnO$_3$ and PrFeO$_3$ decreases with increasing content of dopant (Fe or Mn, respectively). We expect that for small $x$ concentrations only the long range antiferromagnetic ordering of Mn$^{3+}$ ions with a ferromagnetic component exists, Fe ions do not order separately and behave only as a magnetic perturbation effecting magnetic properties of the material. Since the Mn -- O -- Fe magnetic interaction differs from Mn -- O -- Mn interaction the coherent length of long range magnetic ordering becomes smaller with Fe substitution and the ordering temperature decreases. The situation is similar on the opposite site, for $x$ decreasing from 1, where only Fe$^{3+}$ ions order and the ordering temperature decreases due to weaker superexchange magnetic interaction. 

Interesting concentration range, shown in Fig. \ref{fig13} and labeled as Mn + Fe, is the range $0.35 < x < 0.5$ where long range magnetic ordering of both, Mn and Fe ions is present simultaneously. Since the magnetic ordering temperature for Mn ions at concentration $x$ = 0.4 does not fit on the line drawn from region for $x < 0.35$, we estimate that the ordering of Mn ions is induced by already ordered Fe ions. Such an assumption can be underlined by the change in the shape of hysteresis loops around concentration $x$ = 0.35 (Fig. \ref{fig12}). Since the hysteresis loops at temperatures higher than 100 K are linear for all concentrations, this change of shape at low temperatures cannot be connected with the effect of the ordered Fe ions, alone. The interplay of two magnetic sublattices for $x$ = 0.5 can explain the broadening of the hysteresis loop when comparing with PrMnO$_3$ and PrFeO$_3$. However, the hysteresis loop measured at 2 K does not have the shape, which is typical for pure ferromagnetic material neither for polycrystalline sample (Fig. \ref{fig12}b), nor for single crystal (Supplementary data). So we conclude that even at this concentration, both antiferromagnetic and ferromagnetic interactions are present. Our scenario is different from the findings of Ganeshraj et al. \cite{ganeshraj2010}, who reports the ferromagnetic ordering for this concentration.

%
%
\begin{figure}[t]
\begin{center}
\includegraphics[angle=0,width=0.45\textwidth]{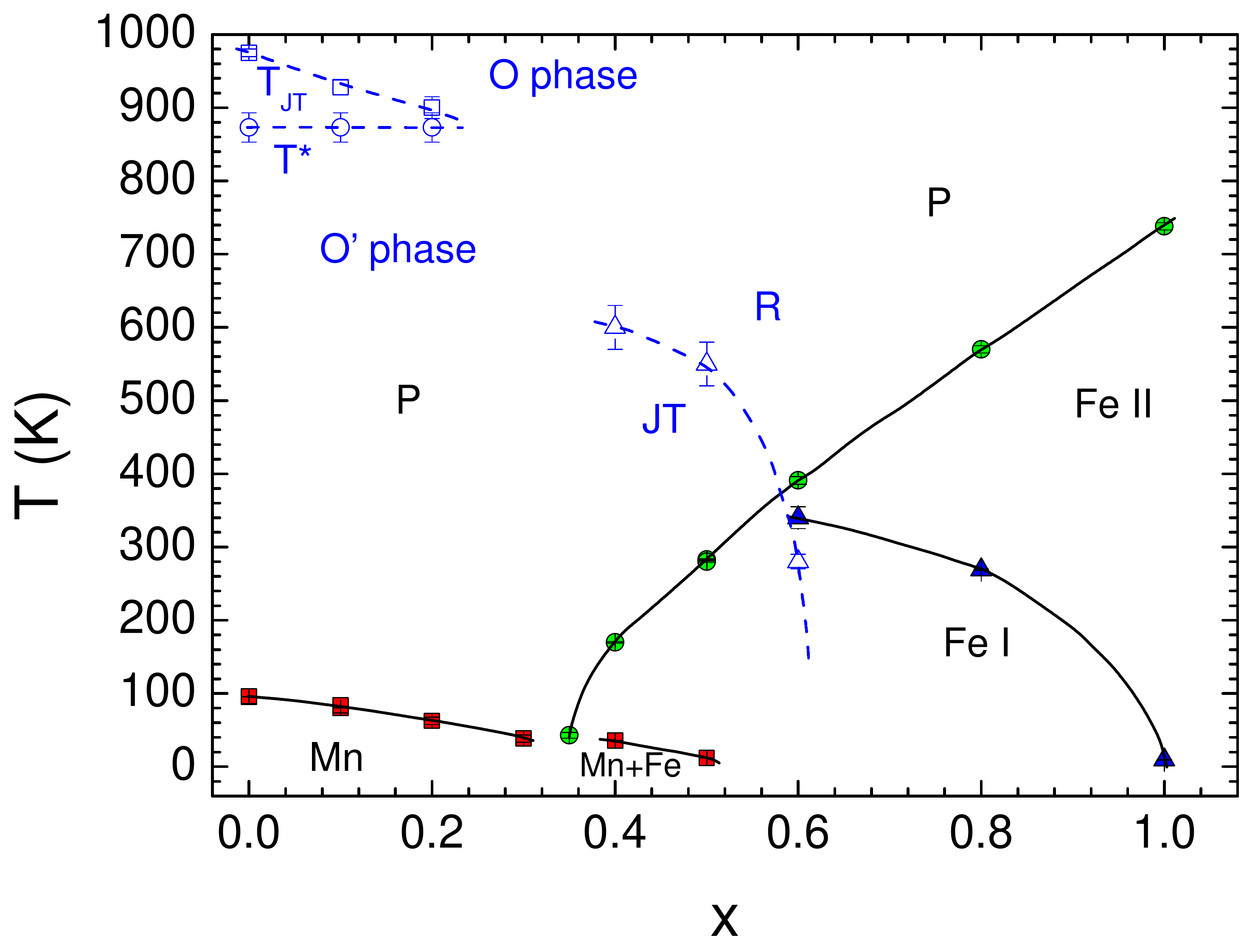}
\pdfcompresslevel=9
\caption{(Color online) The structural (blue open symbols and dashed lines) and magnetic (full symbols and solid lines) phase diagram of PrMn$_{1-x}$Fe$_x$O$_3$ system. O’ and O: two different orthorhombic structures as described in text; JT: orthorhombic distortion is mainly driven by JT distortion of (Mn,Fe)O$_6$ octahedrons; R: orthorhombic distortion is dominated by tilting and rotation of (Mn,Fe)O$_6$ octahedrons; Mn: only Mn sublattice is magnetically ordered; Mn+Fe: both, Mn and Fe ions do magnetically order; FeI and FeII: only Fe sublattice is magnetically ordered, but FeI magnetic structure is different from FeII magnetic structure; P: paramagnetic region. \label{fig13}}
\end{center}
\end{figure}
%
%
  
Additional feature in the magnetic phase diagram is the change of the magnetic structure of iron sublattice for $0.6 \leq x \leq 1$ from Fe I to Fe II (see Fig. \ref{fig13}). The magnetization data suggest that this transition is taking place smoothly in relatively wide temperature range having spin reorientation  character for $x$ = 1. The character of the transition changes to much sharper, spin switching type feature with increasing Mn concentrations. Such a behavior was previously already observed for NdFeO$_3$ \cite{yuan2013}, but also for doped TbMn$_{0.25}$Fe$_{0.75}$O$_3$ \cite{kim2011}, TbMn$_{0.5}$Fe$_{0.5}$O$_3$ \cite{nhalil2015} and DyMn$_{1-x}$Fe$_x$O$_3$ \cite{chiang2011} compounds. Neutron diffraction experiment is highly desirable to study the details of this magnetic phase transition and both, Fe I and Fe II magnetic structures of iron sublattice. 

%
%
\section*{Acknowledgment}
This work was supported by VEGA project 2/0132/16, ERDF EU under the contract No ITMS26220120047 and Slovenian Research Agency (ARRS), Program Number P2-0348.
%
%

\clearpage



\section*{Supplementary material}

# Structural and magnetic study of PrMn$_{1-x}$Fe$_x$O$_3$ compounds – supplementary online material

Matúš Mihalik,[a] Zvonko Jagličić,[b] Magdalena Fitta,[c] Viktor Kavečanský,[a] Kornel Csach,[a] Andrzej Budziak,[c] Jaroslav Briančin,[d] Mária Zentková,[a] Marián Mihalik[a]

[a]Institute of Experimental Physics SAS, Watsonova 47, 040 01 Košice, Slovakia
[b]Institute of Mathematics, Physics and Mechanics and Faculty of Civil and Geodetic Engineering, University of Ljubljana, Slovenia
[c]Institute of Nuclear Physics Polish Academy of Sciences, Radzikowskiego 152, 31-342 Kraków, Poland
[d]Institute of Geotechnics SAS, Watsonova 45, 043 53 Košice, Slovak Republic

## 1. XRPD temperature scans: raw data

PrMnO$_3$:

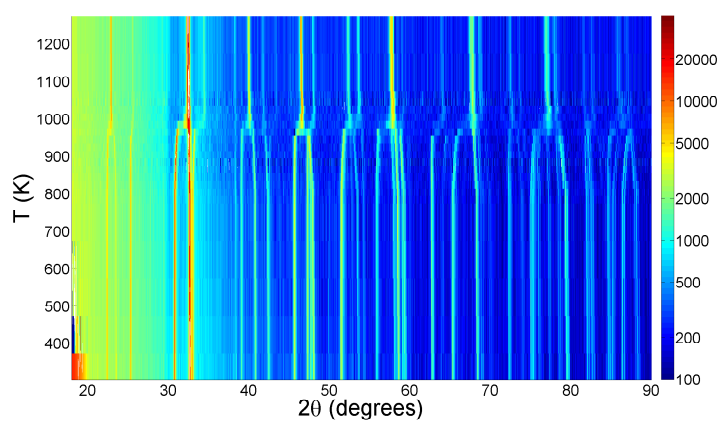

PrMn$_{0.9}$Fe$_{0.1}$O$_3$:

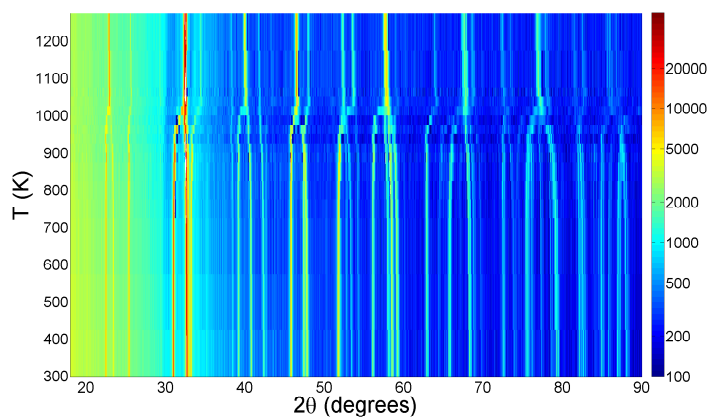

PrMn$_{0.8}$Fe$_{0.2}$O$_3$:

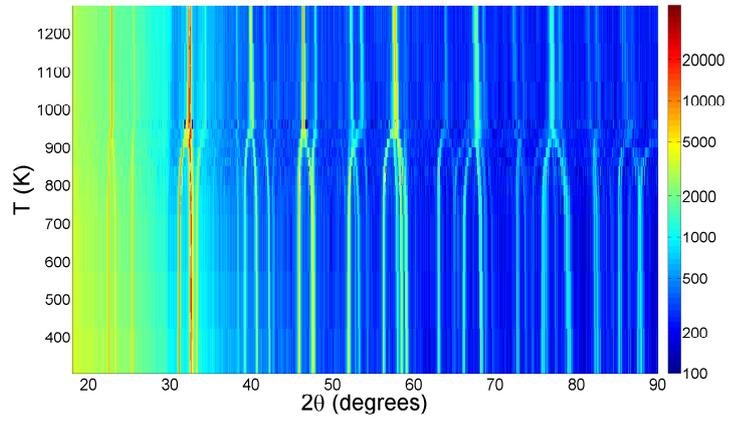

PrMn$_{0.6}$Fe$_{0.4}$O$_3$:

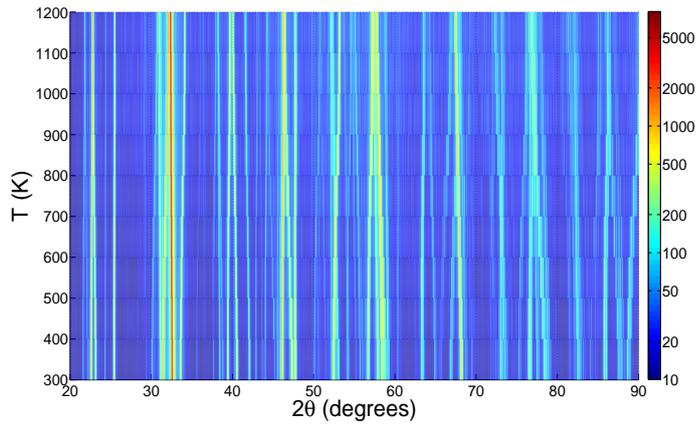

PrMn$_{0.5}$Fe$_{0.5}$O$_3$:

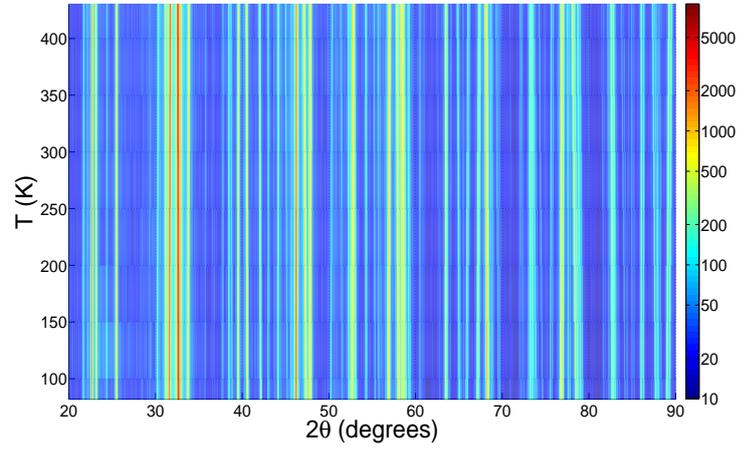

PrMn$_{0.4}$Fe$_{0.6}$O$_3$:

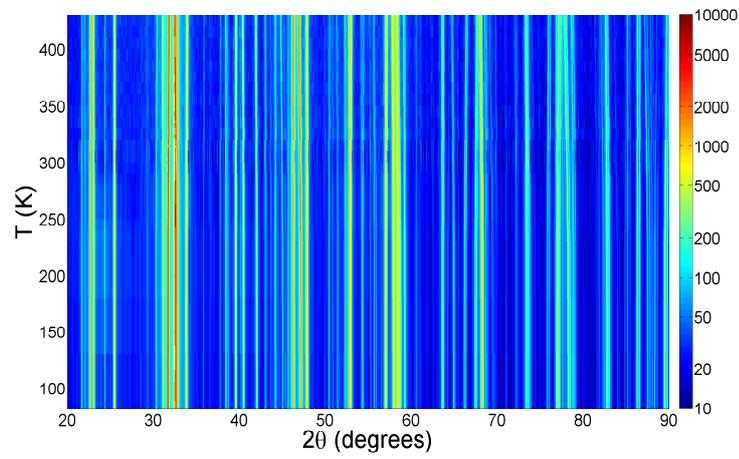

PrMn$_{0.2}$Fe$_{0.2}$O$_3$:

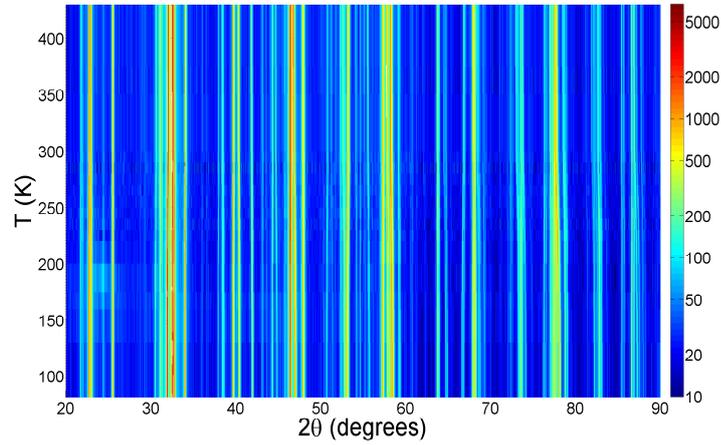

## 2. XRPD temperature scans: analysis

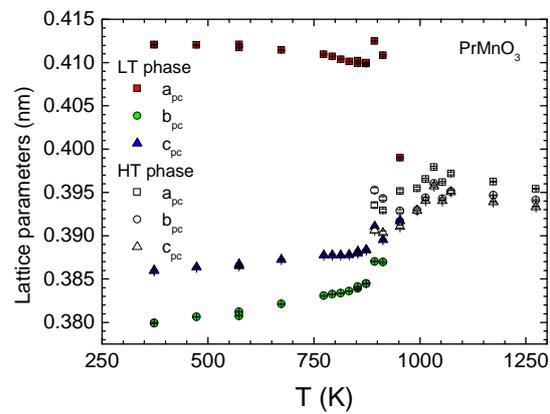

The evolution of the pseudocubic crystallographic parameters for PrMnO3 compound – both crystallographic phases.

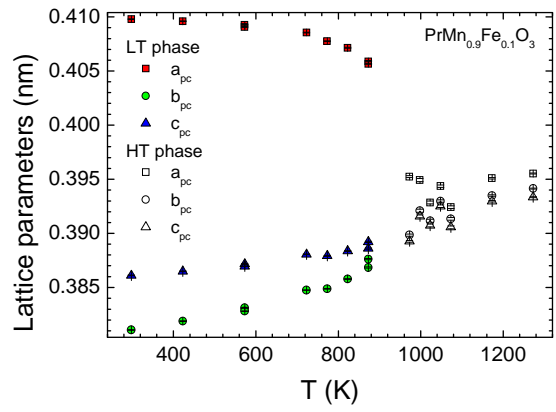

The evolution of the pseudocubic crystallographic parameters for PrMn$_{0.9}$Fe$_{0.1}$O$_3$ compound – both crystallographic phases.

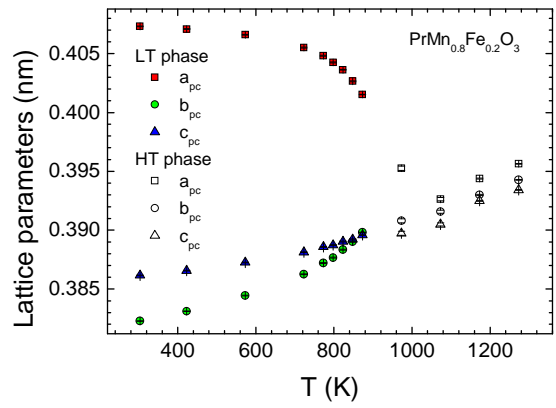

The evolution of the pseudocubic crystallographic parameters for PrMn$_{0.8}$Fe$_{0.2}$O$_3$ compound – both crystallographic phases.

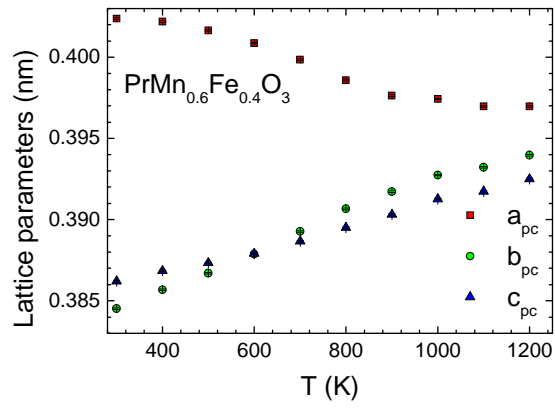

The evolution of the pseudocubic crystallographic parameters for PrMn$_{0.6}$Fe$_{0.4}$O$_3$ compound.

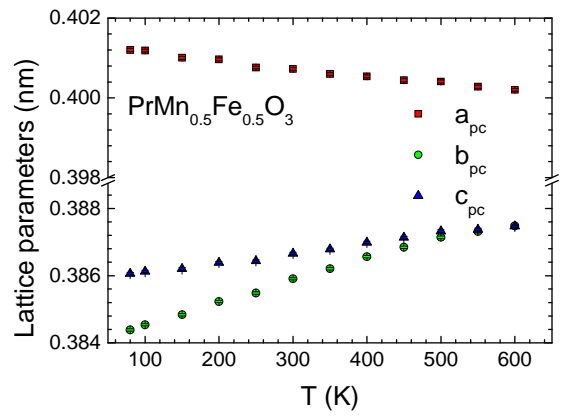

The evolution of the pseudocubic crystallographic parameters for PrMn$_{0.5}$Fe$_{0.5}$O$_3$ compound – crossing between $a_{pc} > c_{pc} > b_{pc}$ and $a_{pc} > b_{pc} > c_{pc}$ regimes in temperature

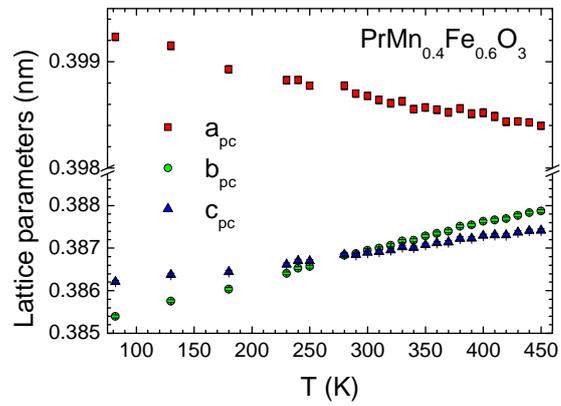

The evolution of the pseudocubic crystallographic parameters for PrMn$_{0.4}$Fe$_{0.6}$O$_3$ compound – crossing between $a_{pc} > c_{pc} > b_{pc}$ and $a_{pc} > b_{pc} > c_{pc}$ regimes in temperature

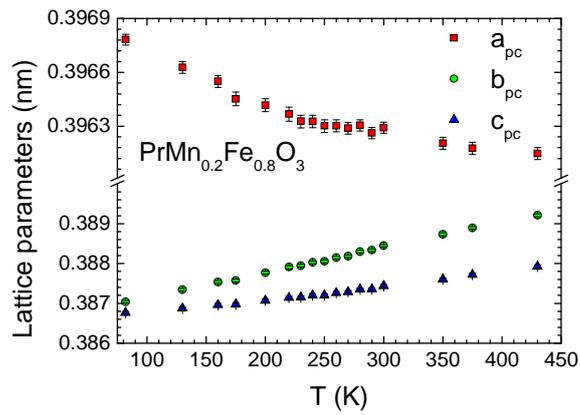

The evolution of the pseudocubic crystallographic parameters for PrMn$_{0.2}$Fe$_{0.8}$O$_3$ compound – crossing between $a_{pc} > c_{pc} > b_{pc}$ and $a_{pc} > b_{pc} > c_{pc}$ regimes in temperature

## 3. ZFC-FC data measured on single crystals

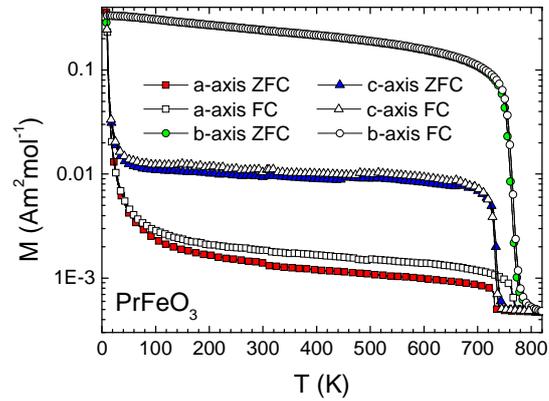

Measurements performed on PrFeO$_3$ compound. Applied magnetic field $\mu_0 H$ = 100 Oe along all three main crystallographic axes.

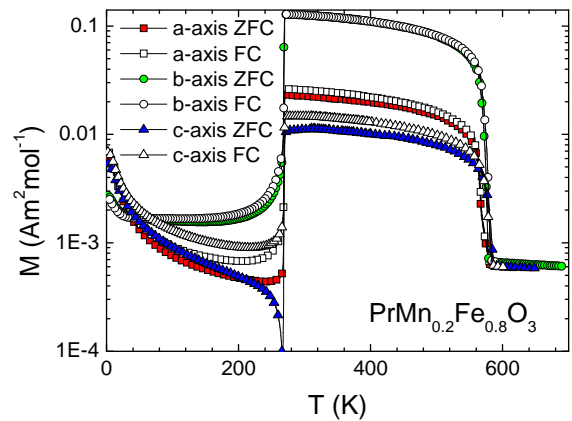

Measurements performed on PrMn$_{0.2}$Fe$_{0.8}$O$_3$ compound. Applied magnetic field $\mu_0 H$ = 100 Oe along all three main crystallographic axes.

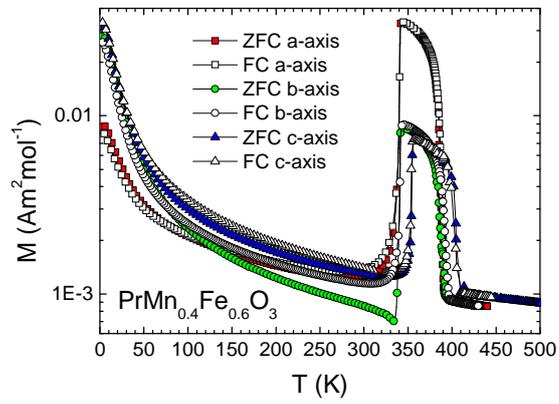

Measurements performed on PrMn$_{0.4}$Fe$_{0.6}$O$_3$ compound. Applied magnetic field $\mu_0 H$ = 100 Oe along all three main crystallographic axes.

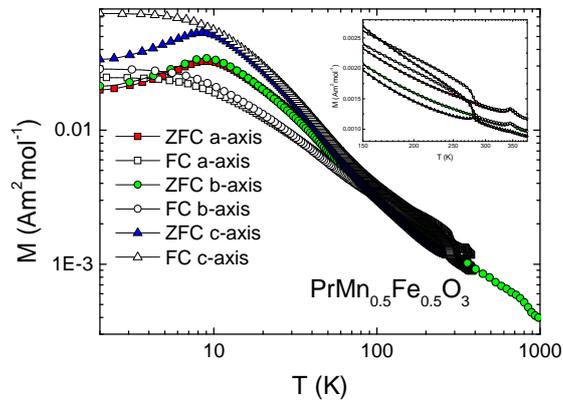

Measurements performed on PrMn$_{0.5}$Fe$_{0.5}$O$_3$ compound. Applied magnetic field $\mu_0 H$ = 100 Oe along all three main crystallographic axes.

## 4. M(B) data measured on single crystals

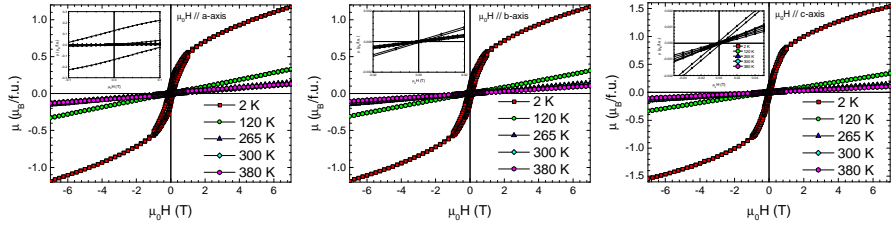

Measurements performed on PrMn$_{0.5}$Fe$_{0.5}$O$_3$ compound with magnetic field applied along all three main crystallographic axes.

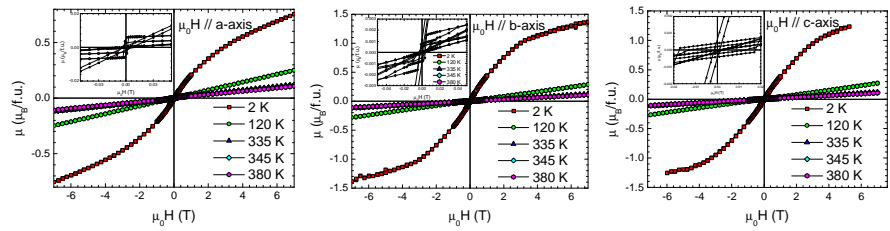

Measurements performed on PrMn$_{0.4}$Fe$_{0.6}$O$_3$ compound with magnetic field applied along all three main crystallographic axes.

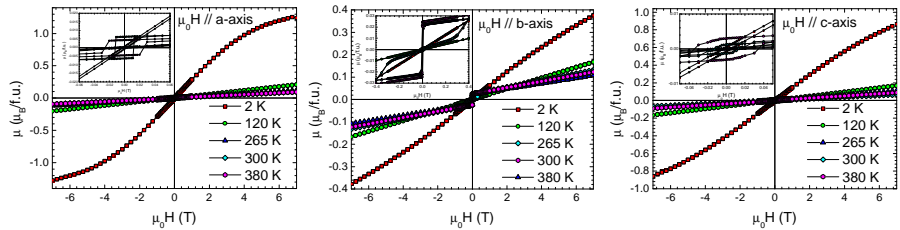

Measurements performed on PrMn$_{0.2}$Fe$_{0.8}$O$_3$ compound with magnetic field applied along all three main crystallographic axes.

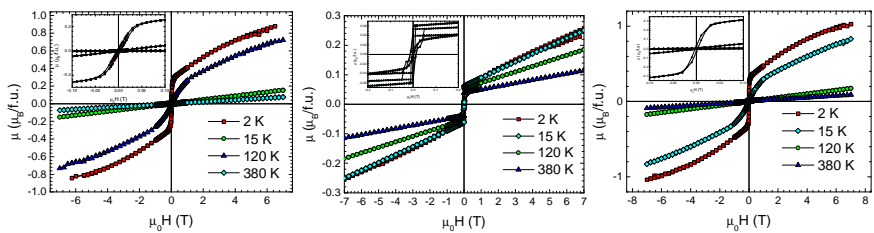

Measurements performed on PrFeO$_3$ compound with magnetic field applied along all three main crystallographic axes.

## 5. Comparison of single crystal and polycrystalline for $PrMn_{0.5}Fe_{0.5}O_3$ compound

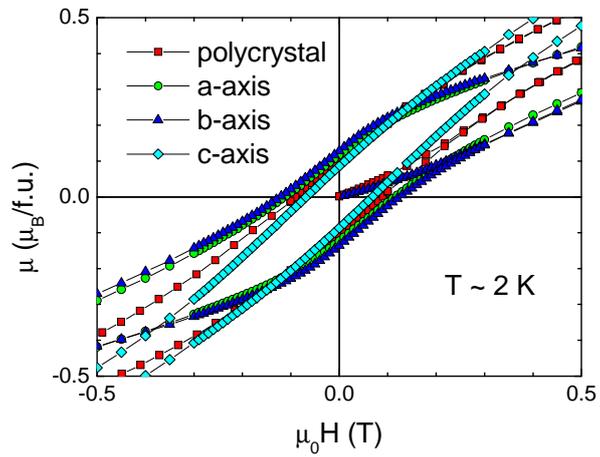

Detail of hysteresis loops measured at temperatures 1.9 K for polycrystal and 2 K for single crystal.

## 6. Curie-Weiss fits for $0 \leq x \leq 0.35$

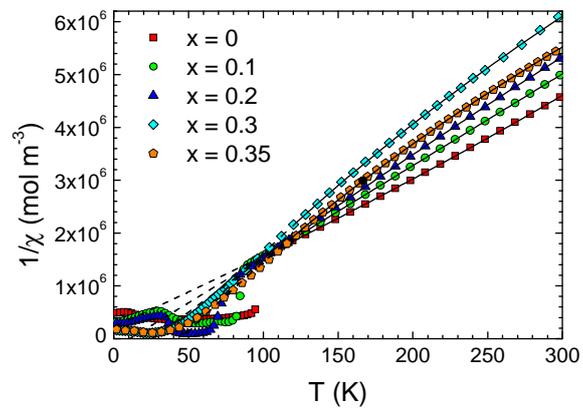

Data obtained on powdered samples with applied magnetic field µ0H = 0.1 T. Full lines represent the best fit with parameters as presented in Table 1; dashed lines are extrapolations.



\end{document}